\documentclass[aps,prc,showpacs,preprint,preprintnumbers,
showkeys,tighten]{revtex4}
\usepackage{amsmath,amssymb,amsfonts}
\usepackage{graphicx}
\usepackage{color}
\allowdisplaybreaks

\begin{document}
\title{Wakes in a Collisional Quark-Gluon Plasma}
\author{Purnendu \surname{Chakraborty}}
\email{purnendu.chakraborty@saha.ac.in}
\author{Munshi Golam \surname{Mustafa}}
\email{musnhigolam.mustafa@saha.ac.in}
\author{Rajarshi \surname{Ray}}
\email{rajarshi.ray@saha.ac.in}
\affiliation{Theory Division, Saha Institute of Nuclear Physics, 1/AF
  Bidhannagar, Kolkata 700064.}
\author{Markus H. \surname{Thoma}}
\email{thoma@mpe.mpg.de}
\affiliation{Max-Planck-Institut
f\"ur extraterrestrische Physik, P.O. Box 1312, 85741 Garching, Germany} 
\begin{abstract}
Wakes created by a parton moving through a static and infinitely extended  
quark-gluon plasma are considered. In contrast to former investigations 
collisions within the quark-gluon plasma are taken into account using a
transport theoretical approach (Boltzmann equation) with a Bhatnagar-Gross-Krook
(BGK) collision term. Within this model it is shown that the wake structure
changes significantly compared to the collisionless case. 
\end{abstract}
\maketitle
In ultrarelativistic heavy-ion experiments (SPS, RHIC) indications 
have been found for a temporary formation of a strongly coupled
quark-gluon plasma (QGP)
phase in the fireball created in nucleus-nucleus collisions (see e.g. 
Ref.\cite{Ullrich2007}). The main problem of this programme is to find
clear signatures for the presence of a short-living QGP phase in this 
fireball. One promising class of signatures are hard probes, in particular
high-energy partons of a few GeV or more, i.e. with an energy much higher 
than the temperature of the QGP. Energetic partons with a large 
transverse momentum, produced in initial hard 
parton collisions before the formation of an equilibrated fireball, 
propagate through the fireball. In this process they will lose energy,
which depends on the state of the fireball (QPG or hadronic matter).
Energetic partons manifest themselves at jets arriving in the detectors.
The amount of jet quenching due to the propagation of the leading parton
of the jet through the fireball can therefore serve as a signature for the 
QGP formation \cite{Pluemer1990,Baier2000}. A strong jet quenching and 
suppression of high $p_T$ hadron spectra have been observed at RHIC indicating
the presence of a QGP phase \cite{RHIC}. Another phenomenon related to the 
propagation of a high-energy parton through the QGP are wakes and Mach cones
which may be observable as conical flow and shock waves in particle correlations 
\cite{Stoecker2005,Ruppert2005,Satarov2005,Shuryak2006,Chakraborty2006,remark1}.

In electromagnetic plasmas wakes are a well known phenomena. Theoretically 
they can be investigated by considering transport equations. In the 
collisionless case they follow from the dielectric functions by
solving the Vlasov equation together with the Poisson equation. 
It has been shown in this way that wakes
are also created in a QGP, leading eventually to attractive potentials 
and Mach cones as well known in plasma physics (see below). 
Attractive potentials between partons might
lead to bound states and may modify the $J/\psi$ suppression pattern 
\cite{Chakraborty2006}. In the static case, i.e. for a parton at 
rest, the wake potential reduces to the usual Yukawa potential
describing Debye screening \cite{Mustafa2005}. 

Here we want to extend these investigations by taking into account collisions 
in the QGP. For this purpose we start from the Boltzmann equation
approximating the collision term by the BGK description \cite{Bhatnagar1954}, allowing
an analytic expression for the dielectric functions \cite{Carrington2004,remark2}. In this
approach the collision term is replaced by a momentum independent collision rate 
which we take as a parameter. This model has been used to study dispersion
relations \cite{Carrington2004} and instabilities \cite{Schenke2006} in a
collisional QGP. Whereas the changes in the case of the dispersion relations
are marginal - however the longitudinal plasma modes cross the light cone in a 
collisional plasma -, 
instabilities can be suppressed efficiently due to collisions. In the case of 
complex plasmas, i.e., low-temperature discharge plasmas with microparticles such
as dust grains \cite{Fortov2005}, wake potentials due to ion flow, leading
to an attractive force between the negatively charged microparticles, are investigated
intensively (see e.g. \cite{Bashkirov2004}). Also collisions between the ions and
the neutral gas have been taken into account in this case and
have been shown to truncate the wakes effectively \cite{Lampe2000}. In general 
the collisions modify the wake potential by decelerating the ions.
Therefore the positive ion cloud, responsible for the attractive part of the wake 
potential, is concentrated nearer to the dust grain, causing a narrower but deeper
potential well.

The longitudinal and transverse dielectric functions of a collisional QGP
within the BGK approach are given by~\cite{Carrington2004}, 
\begin{eqnarray}
\epsilon_l\left(\omega,k\right) &=& 1 + \frac{m_D^2}{k^2}\left(
1 - \frac{\omega + i\nu}{2k}\ln{\frac{\omega + i\nu +k}{\omega + i\nu -k}}
\right)\left(
1 - \frac{i\nu}{2k}\ln{\frac{\omega + i\nu +k}{\omega + i\nu -k}}
\right)^{-1}\nonumber\\
\epsilon_t\left(\omega,k\right) &=& 1 - \frac{m_D^2}{2\omega\left(\omega +
    i\nu\right)} \left\{1 + \left[\frac{\left(\omega + i\nu\right)^2}{k^2}
-1 \right]\left(
1 - \frac{\omega + i\nu}{2k}\ln{\frac{\omega + i\nu +k}{\omega + i\nu -k}}
\right) \right\}\,,\label{e_l.and.e_t}
\end{eqnarray}
where $\nu$ is the collision rate and $m_D = \sqrt{1+n_f/6}\, gT$ the Debye screening mass. 
Here $n_f$ is the number of quark flavors in the QGP. From the dielectric functions the dispersion
relations follow \cite{Carrington2004}, e.g. the longitudinal mode (plasmon) from 
$\epsilon_l(\omega,k)=0$. Note that $\omega $ has a real and imaginary part here. 

The induced charge density by a parton with color charge $Q^a$ propagating 
through the QGP with velocity $v$ follows from \cite{Chakraborty2006}
\begin{equation}
\rho^a_{\rm{ind}}(t,{\vec{\mathbf r}})= 2\pi Q^a \
\int \frac{d^3k}{(2\pi)^3} \int \frac{d\omega}{2\pi} 
\exp\left [ i\left ({\vec{ k}} \cdot {\vec { r}} - \omega t
\right )\right ] 
\left [\frac{1}{\epsilon_l(\omega,k)}-1 \right ]\ \delta(\omega -{\vec
{k}} \cdot {\vec { v}}) \, \, 
. \label{indrho2}
\end{equation}
Due to the delta function $\omega$ is restricted here to real values and 
is always in the space-like region for $0<v<1$.

The induced charge density  using cylindrical coordinates
can be written as \cite{remark3} 
\begin{eqnarray}
\rho^a_{\rm{ind}}(\rho,z,t) &=& \frac{Q^a}{2\pi^2v} \int_0^\infty d\kappa \, \kappa 
\, J_0 (\kappa \rho) \int_{0}^{\infty}\ d\omega \, 
\Biggl \{\cos \left [ \omega \left (\frac{z}{v}-t\right )\ \right] 
\left ( \frac{{\rm Re}\,\epsilon_l}{\Delta} - 1 \right )\nonumber \\ 
&+& \sin \left [\omega \left (\frac{z}{v}-t\right ) \right ]
\frac{{\rm Im}\,\epsilon_l}{\Delta}  
\Biggr \} \, \, , \label{indrho4}
\end{eqnarray}
where $J_0$ is the Bessel function, 
$\Delta=[{\rm Re}\,\epsilon_l]^2+ [{\rm Im}\,\epsilon_l]^2$, and 
$\kappa=\sqrt{k^2-{\omega^2}/{v^2}}$.

The real and imaginary parts of the dielectric functions read
\begin{eqnarray}
{\rm Re}\,\epsilon_l &=& 1+ \frac{m_D^2}
{4k^4f\left(\omega,k,\nu\right)}\left[
4k^2 + \nu^2\left(\theta^2 + \ln^2{R}\right) -2k\left(\omega\ln{R} 
+2\nu\theta\right)
\right]\,,\nonumber\\
{\rm Im}\,\epsilon_l &=& \frac{m_D^2 \omega}
{4k^4f\left(\omega,k,\nu\right)}\left[
2k\theta -\nu\left(\ln^2{R}+\theta^2\right)
\right] \,,\nonumber\\
\text{where}&&\nonumber\\
R&=& \frac{\sqrt{\left(\omega^2 -k^2 +\nu^2\right)^2
+4k^2\nu^2}}{\left(\omega - k\right)^2 + \nu^2}\nonumber\,,\\
\theta &=& \arccos \frac{\omega^2-k^2 +\nu^2}
{\sqrt{\left(\omega^2 -k^2 +\nu^2\right)^2+4k^2\nu^2}}\, ,\\
\text{and}&&\nonumber\\
f\left(\omega,k,\nu\right) &=& \left(1 - \frac{\nu\theta}{2k}\right)^2+
\frac{\nu^2\ln^2{R}}{4k^2}\,.\label{re_im_el}
\end{eqnarray} 

Solving (\ref{indrho4}) together with (\ref{re_im_el}) numerically \cite{remark4} leads to
the induced charge densities shown in Figs.1 and 2. The induced densities are 
proportional to $m_D^3$ and are scaled with $m_D$ as well as with the color charge 
$Q^a$ of the moving  parton. Assuming for example $n_f=2$, $g=2$ (corresponding to 
$\alpha_s = 1/\pi$), and $T=250$ MeV, we get a typical value of $m_D = 580$ MeV.
In Fig.1 the parton 
velocity is $v=0.55\> c$ and $v=0.99\> c$ in Fig.2. The left panels show 
the 3 dimensional plots for the choice $\nu = 0$ to $\nu = 0.8 m_D$ 
\cite{remark5}. The right panels show the contour plots of the induced
charge density. One observes a clear modification of the 
induced charge density due to the collision rate. The general structure of the wakes,
corresponding to a modified screening, a cone-like structure, and wave excitation 
has been discussed in detail in Ref.\cite{Chakraborty2006} in the collisionless case. 
However, in the case of collisions these structures are smeared out
and less pronounced for increasing $\nu$. We interpret this as the fact that the collisions,
leading to thermalization, reduce the anisotropy caused by the perturbation of the 
moving charge in the QGP.  

The wake potential is given by (see appendix)
\begin{equation}
\label{wake_potential1}
\Phi^a ({\vec r}, t; {\vec v}) = \frac{Q^a}{2 \pi^2} 
\int d^3k \frac{e^{-i{\vec k}\cdot ({\vec r}-{\vec v}t)}}{k^2 \epsilon _l 
(\omega = {\vec v}\cdot {\vec k},k)}.
\end{equation}
It is of the same form as in the collisionless case, where it can be shown that
in cylindrical coordinates it reduces to \cite{Chakraborty2006}
\begin{eqnarray}
\Phi^a\left(\rho,z,t\right) &=& \frac{2Q^a}{\pi v}\int_0^\infty
d\kappa \, \kappa  \, J_0\left(\kappa\rho\right) \int_0^\infty\, d\omega
\frac{1}{k^2 \Delta} \Biggl \{\cos \left[{\omega\left (\frac{z}{v}
-t\right)}\right]\,{\rm{Re}} \, \epsilon_l \nonumber \\
&& + \sin \left[\omega\left(\frac{z}{v}-t\right)\right]\,{\rm{Im}}
\,\epsilon_l \Biggr \}\, . \label{potcyl}
\end{eqnarray}

In Figs.3 and 4 the wake potentials scaled with respect to $m_D$ 
are shown for a parton with a given color charge $Q^a$ for velocities 
$v=0.55\> c$ and $v=0.99\> c$ and various values for the collision rate from
$\nu = 0$ to $\nu = 0.8 m_D$. It can be seen that collisions modify the strength and
the structure of the wake potential significantly. Again, as in the case of the induced
charge density the potential structure is washed out due to the collisions between the 
partons in the QGP. However, the potential becomes also stronger at finite collision rates,
indicating that collisions in the QGP can enhance the attractive and repulsive parts of 
the interaction potential of a fast parton as in the case of non-relativistic plasmas.
In particular, the attractive 
part behind the particle becomes deeper which might lead to an even stronger
binding of diquark states propagating through the QGP 
as discussed in Ref.\cite{Chakraborty2006}. (Note that the induced charge density
and wake potential diverge for point-like particles at the origin as in the static
case corresponding to a Yukawa potential.)
 
In conclusion, we have applied the BGK approach in order to study the influence of
collisions on the wake formation in the QGP. Clearly, this approach allows only 
qualitative statements. A consistent QCD approach, in which for example
the collision rate are momentum dependent, is beyond the scope of this work.
Keeping this in mind, we regard the (momentum independent) collision rate in
our model as a free parameter. Anyway, our model demonstrates clearly
that collisions in the QGP lead to a significant modification of
wakes created by fast partons compared to the collisionless case. In particular, the
wake structure is less pronounced. Hence possible observable effects such as the conical flow 
and wave excitations will be reduced. On the other hand, the attractive potential well
becomes deeper. Similar effects of collisions have been observed in electromagnetic 
plasmas.  

\bigskip

{\bf Acknowledgments}
We would like to thank M. Strickland for helpful discussions.
M.H.T. would like to thank A. Ivlev for information on wakes in complex plasmas.
MGM and RR are also thankful to S. Mallik for useful discussions.

\section{Appendix}

Here we will explicitly derive equation (\ref{wake_potential1}) for the wake potential in the case of
collisions. For the collisionless case it follows from a combination of the
linearized Vlasov equation and the Poisson equation. Here we start from 
the kinetic equation with a BGK-collision term. The derivation from the
equilibrium distribution function $f_0(p)$ is then given by (see (14) in 
Ref.\cite{Carrington2004})
\begin{equation}
\label{deltaf}
\delta f(\vec p,\vec r,t) = \left[-i q \vec E \cdot \frac{\partial f_0(p)}{\partial \vec p}
+i\nu \eta f_0(p)\right] D^{-1},
\end{equation}
where $q$ is the charge of the plasma particles and  
$D:= \omega +i\nu-\vec k \cdot \vec v$ with $\vec v= {\vec p}/p$
in the ultrarelativistic case ($v = 1$).

The quantity $\eta$ is given by
\begin{equation}
\label{eta}
\eta = \frac{-i q}{N_0} \int dp\, \frac{(\vec E\cdot \vec p)}{p} 
\frac{\partial f_0(p)}{\partial p}D^{-1}
\left[1-\frac{i\nu}{N_0}\int dp\, f_0(p)D^{-1}\right]^{-1} 
\end{equation}
with the equilibrium particle number density $N_0$ and $dp=d^3p/(2\pi )^3$.

The Poisson equation for the potential $\Phi$ of a point charge $Q$ with velocity $\vec v$,
where we drop the color index $a$ for convenience,
in the plasma reads   
\begin{equation}
\label{Poisson1}
\Delta \Phi ({\vec r}, t; {\vec v}) = -4\pi Q\, \delta ({\vec r} -{\vec v}t) - \rho ({\vec r},t),
\end{equation}
where $\rho = q\int dp \, \delta f$  is the induced charge density in the plasma.
In momentum space it reads
\begin{equation}
\label{Poisson2}
k^2 \Phi (\omega, {\vec k}, {\vec p}) = 8\pi^2 Q\, \delta (\omega -{\vec v}\cdot {\vec k}) + 
q\int dp\, \delta f (\omega, {\vec k},{\vec p}).
\end{equation}
Using ${\vec E} = -i {\vec k} \Phi$ in momentum space and $\partial f_0(p)/\partial {\vec p}
= {\vec v}\partial f_0(p)/\partial p$ we can write the Poisson equation by combining
(\ref{Poisson2}) with (\ref{deltaf}) and (\ref{eta}) as
\begin{equation}
\label{Poisson3}
k^2 \Phi (\omega, {\vec k}, {\vec p}) = 8\pi^2 Q\, \delta (\omega -{\vec v}\cdot {\vec k}) + 
k^2 (1-\chi) \Phi,
\end{equation}
where
\begin{equation}
\label{kappa}
\chi (\omega ,k) = 1+ \frac{q^2}{k^2} \int dp\, (\vec v\cdot \vec k) 
\frac{\partial f_0}{\partial p}D^{-1} - \frac{i\nu q \eta}{k^2\Phi }\int dp\, f_0 D^{-1}. 
\end{equation}
Solving (\ref{Poisson3}) for $\Phi$ and Fourier transforming it we obtain the wake potential
\begin{equation}
\label{wake_potential2}
\Phi ({\vec r}, t; {\vec v}) = \frac{Q}{2 \pi^2} 
\int d^3k\, \frac{e^{-i{\vec k}({\vec r}-{\vec v}t)}}{k^2 \chi
(\omega = {\vec v}\cdot {\vec k})}.
\end{equation}
Comparing $\chi $ with the longitudinal dielectric function in the BGK approximation
(see (17) and (18) in \cite{Carrington2004}),
\begin{eqnarray}
\label{dielectric}
\epsilon _l (\omega ,k) &&= 1 + \frac{q^2}{\omega k^2}
\int dp\, (\vec v\cdot \vec k)^2 
\frac{\partial f_0}{\partial p}D^{-1} - \frac{i\nu q \eta}{\omega k^2\Phi }\int dp \,
(\vec v\cdot \vec k) f_0 D^{-1}
\end{eqnarray}
we find $\chi (\omega = {\vec v}\cdot {\vec k}) = \epsilon _l 
(\omega = {\vec v}\cdot {\vec k})$. Hence 
(\ref{wake_potential2}) coincides with (\ref{wake_potential1}) and has the same form
as in the collisionless case \cite{Chakraborty2006}.  

This result can also be derived more generally, i.e. in an homogeneous and isotropic
plasma with and without collisions, assuming linear response theory for the induced
charge density \cite{Chakraborty2006}, $\rho_{ind} = (1/\epsilon_l -1) \rho_{ext}$,
where the total charge density $\rho_{tot}=\rho_{ext}+\rho_{ind}$,
and using the Poisson equation $\Phi = 4\pi \rho_{ext}/(k^2 \epsilon_l)$ with
$\rho_{ext}=2\pi Q \delta(\omega-{\vec v}\cdot{\vec k})$ after Fourier transformation.

\begin{figure}[!htbp]
\begin{minipage}[t]{8cm}
\includegraphics[width=8cm,keepaspectratio]{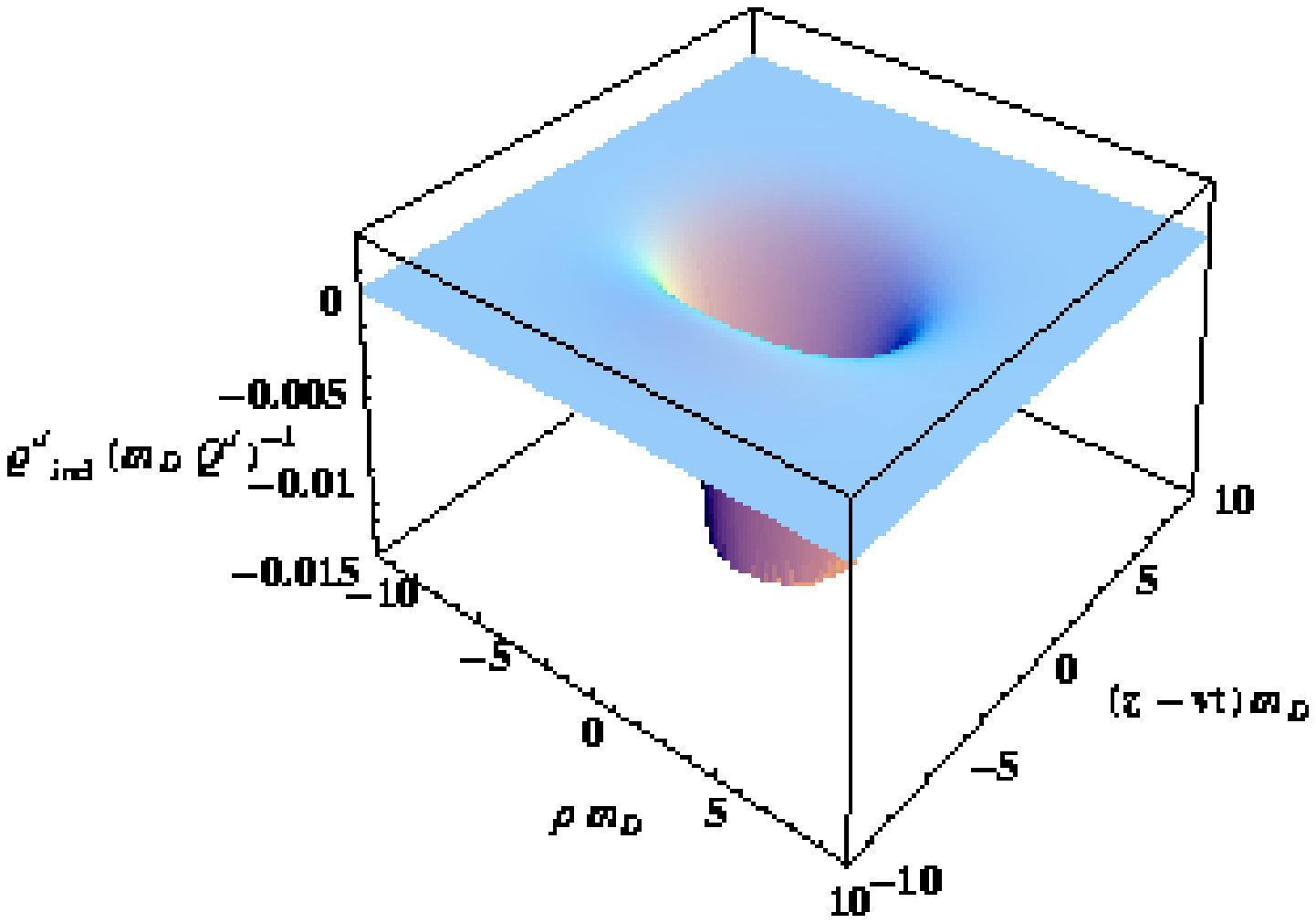}
\end{minipage}
\hfill
\begin{minipage}[t]{8cm}
\includegraphics[width=8cm,keepaspectratio]{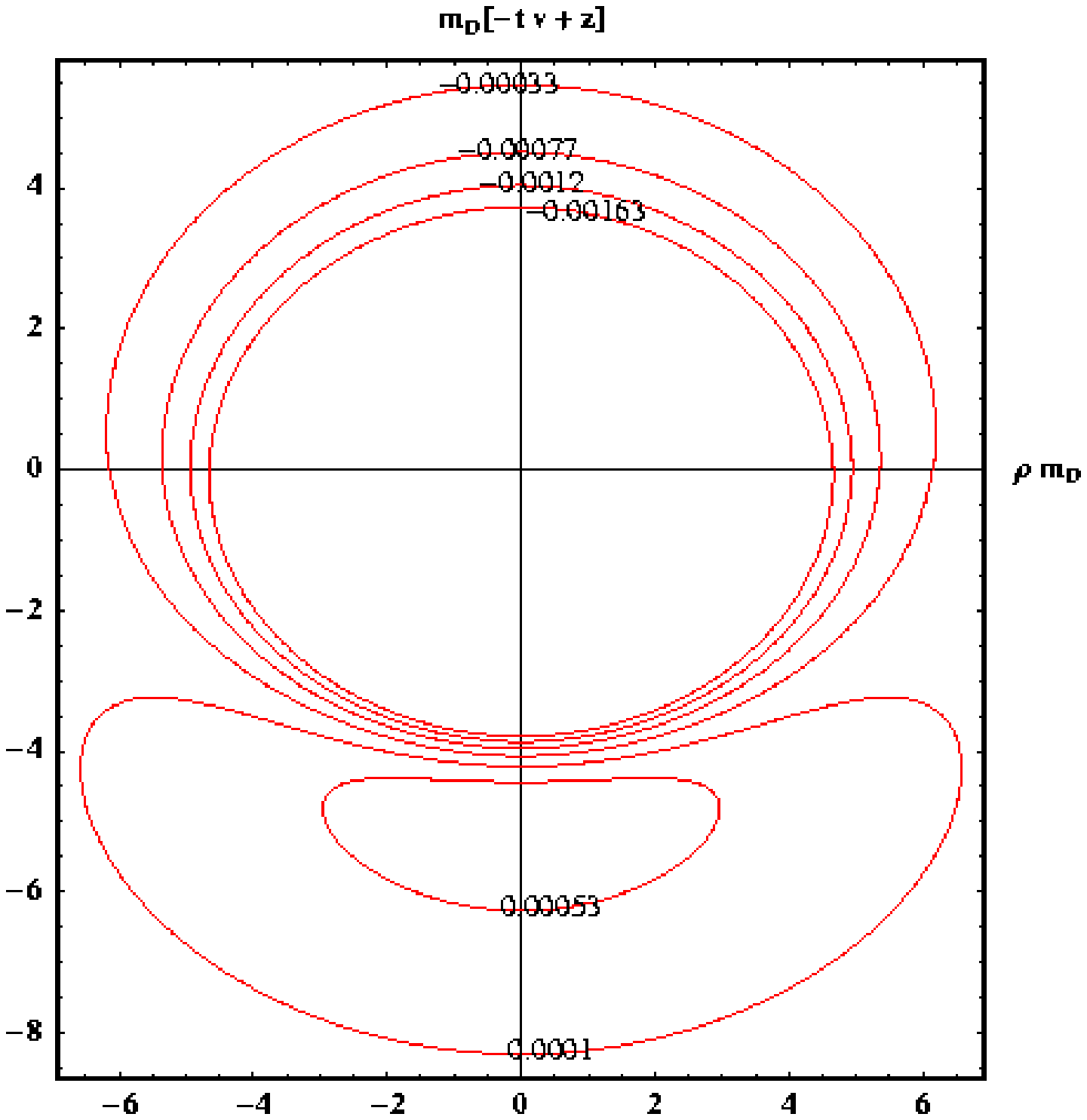}
\end{minipage}
\hfill
\begin{minipage}[t]{8cm}
\includegraphics[width=8cm,keepaspectratio]{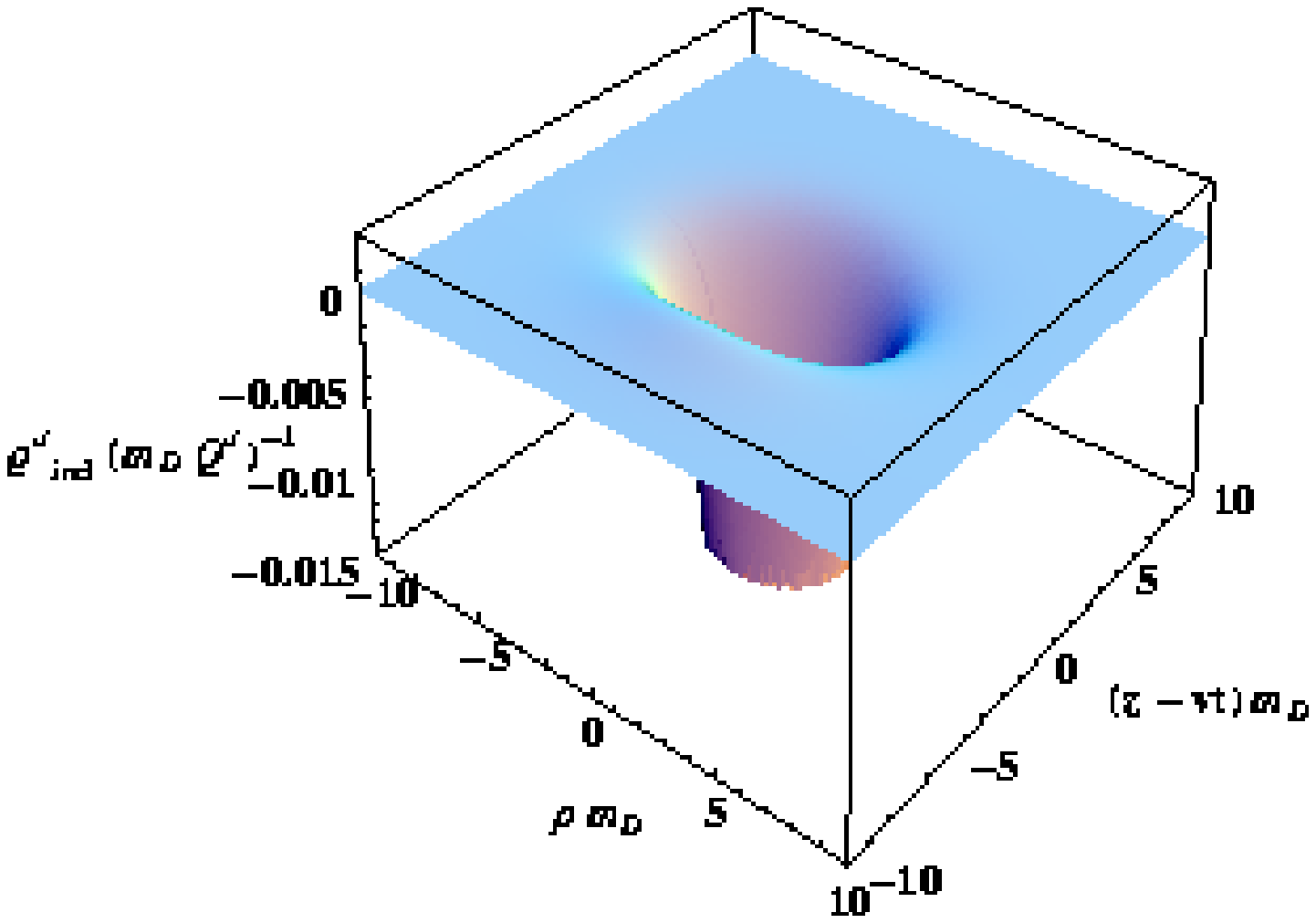}
\end{minipage}
\hfill
\begin{minipage}[t]{8cm}
\includegraphics[width=8cm,keepaspectratio]{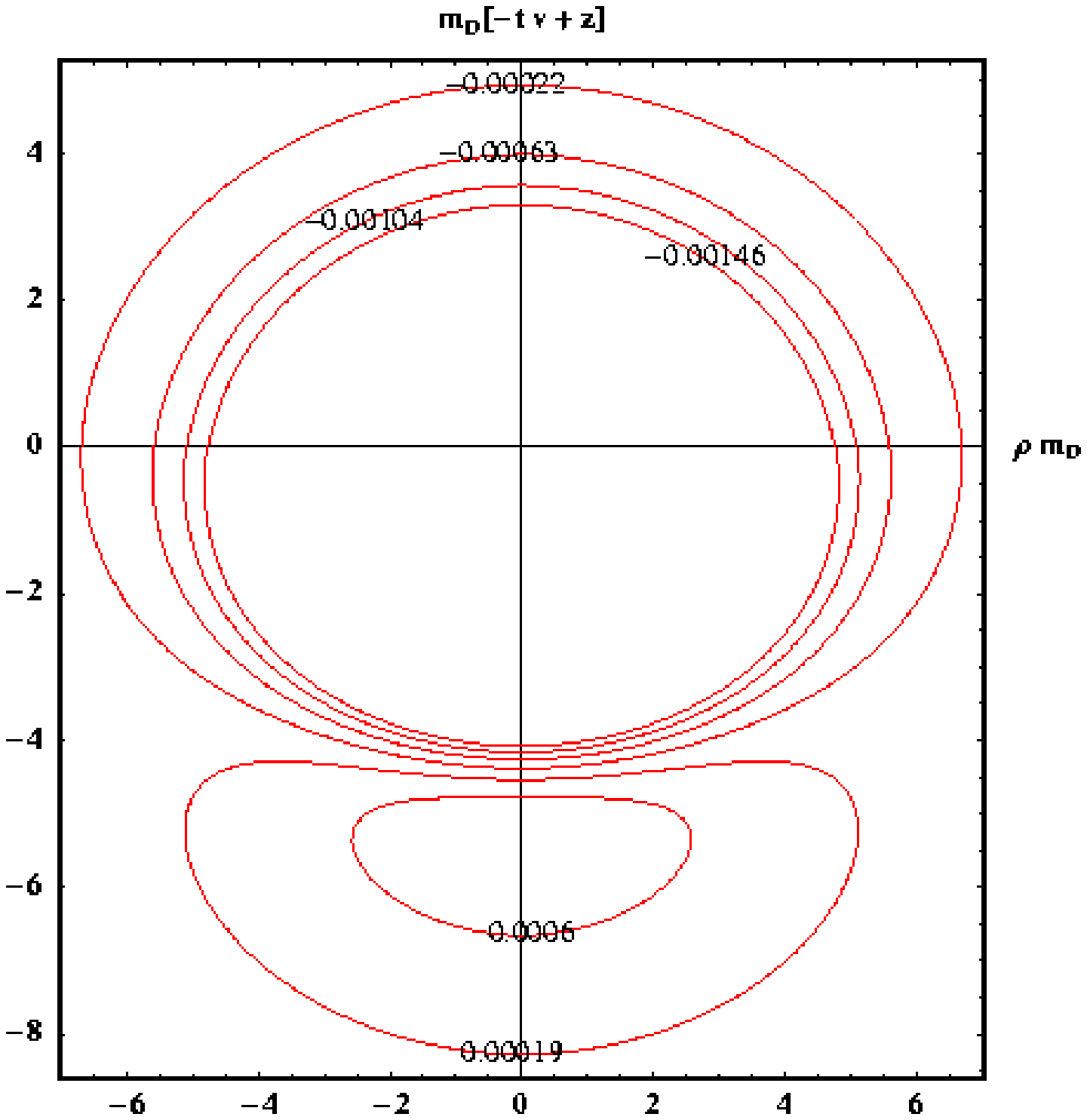}
\end{minipage}
\hfill
\begin{minipage}[t]{8cm}
\includegraphics[width=8cm,keepaspectratio]{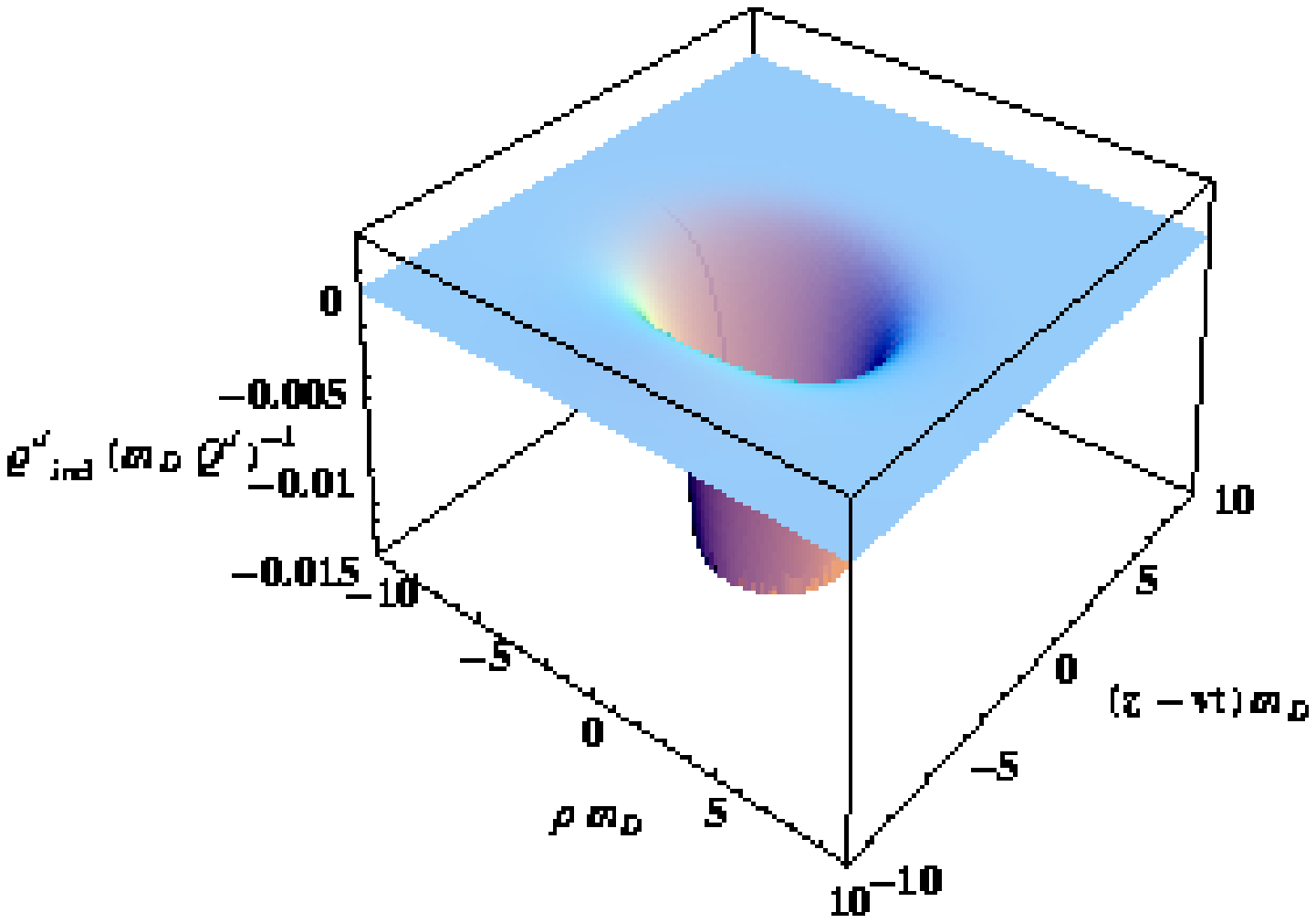}
\end{minipage}
\hfill
\begin{minipage}[t]{8cm}
\includegraphics[width=8cm,keepaspectratio]{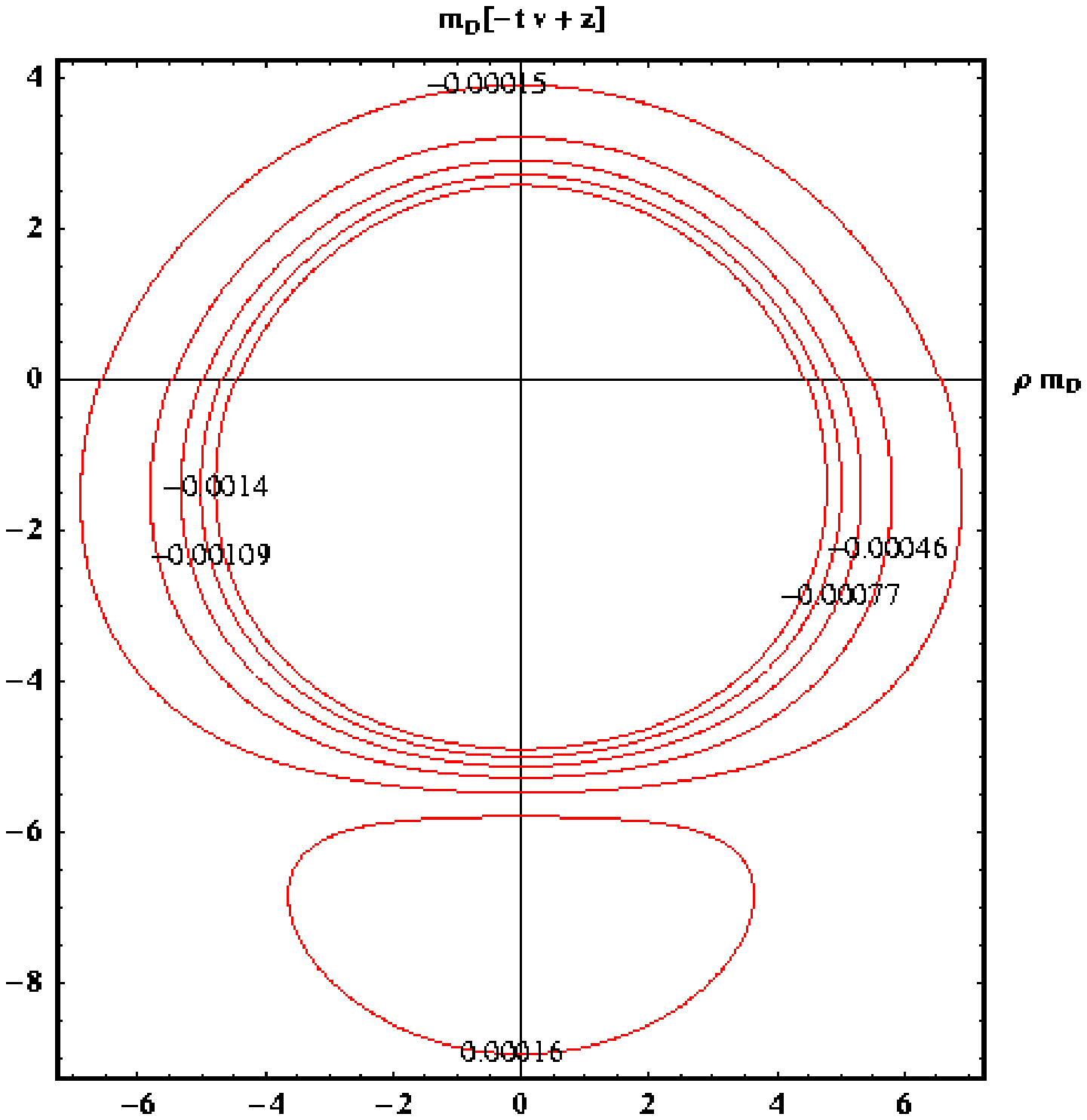}
\end{minipage}
\end{figure}

\begin{figure}[!htbp]
\begin{minipage}[t]{8cm}
\includegraphics[width=8cm,keepaspectratio]{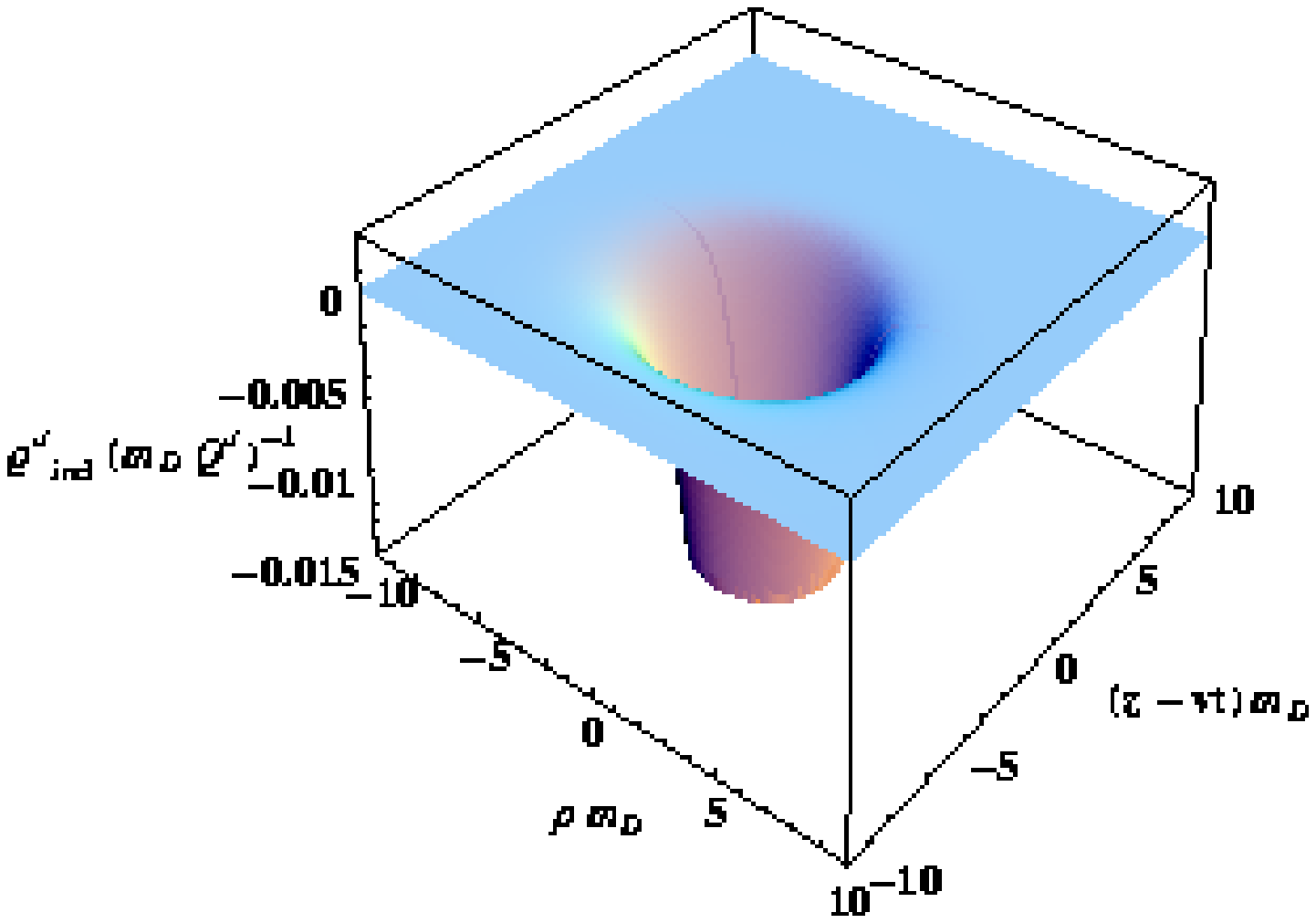}
\end{minipage}
\hfill
\begin{minipage}[t]{8cm}
\includegraphics[width=8cm,keepaspectratio]{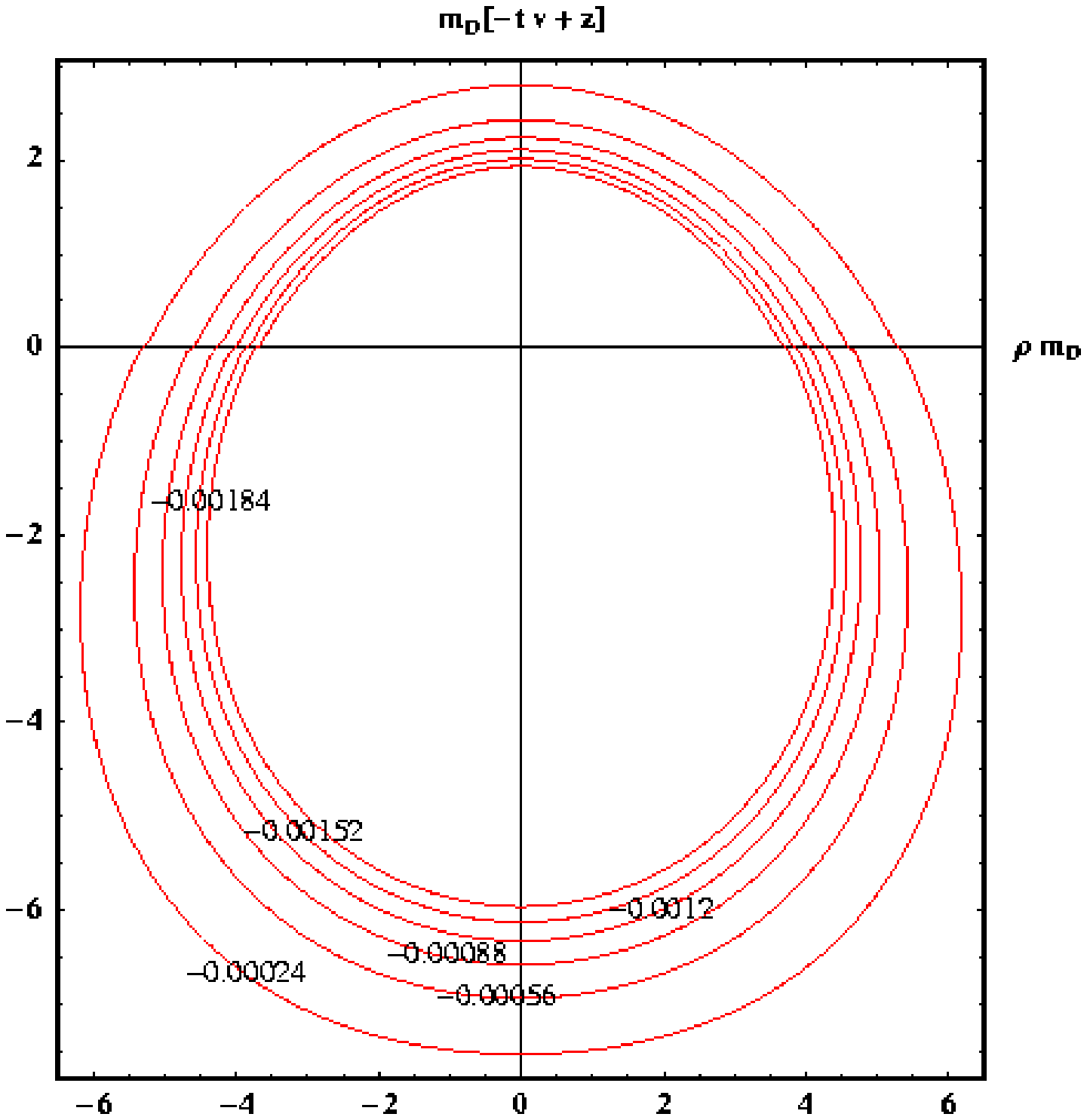}
\end{minipage}
\caption{\label{ichargedendity_v55} Left panel: Spatial distribution of 
the induced charge density for a parton moving with velocity $v=.55c$ 
with collisions for $(\nu=0, \, \,  0.2m_{D}, \, \, 0.5m_{D}, 0.8m_{D})$ 
(Top to Bottom).
Right panel: These plots show the corresponding contour plots}
\end{figure}

\newpage

\begin{figure}[!htbp]
\begin{minipage}[t]{8cm}
\includegraphics[width=8cm,keepaspectratio]{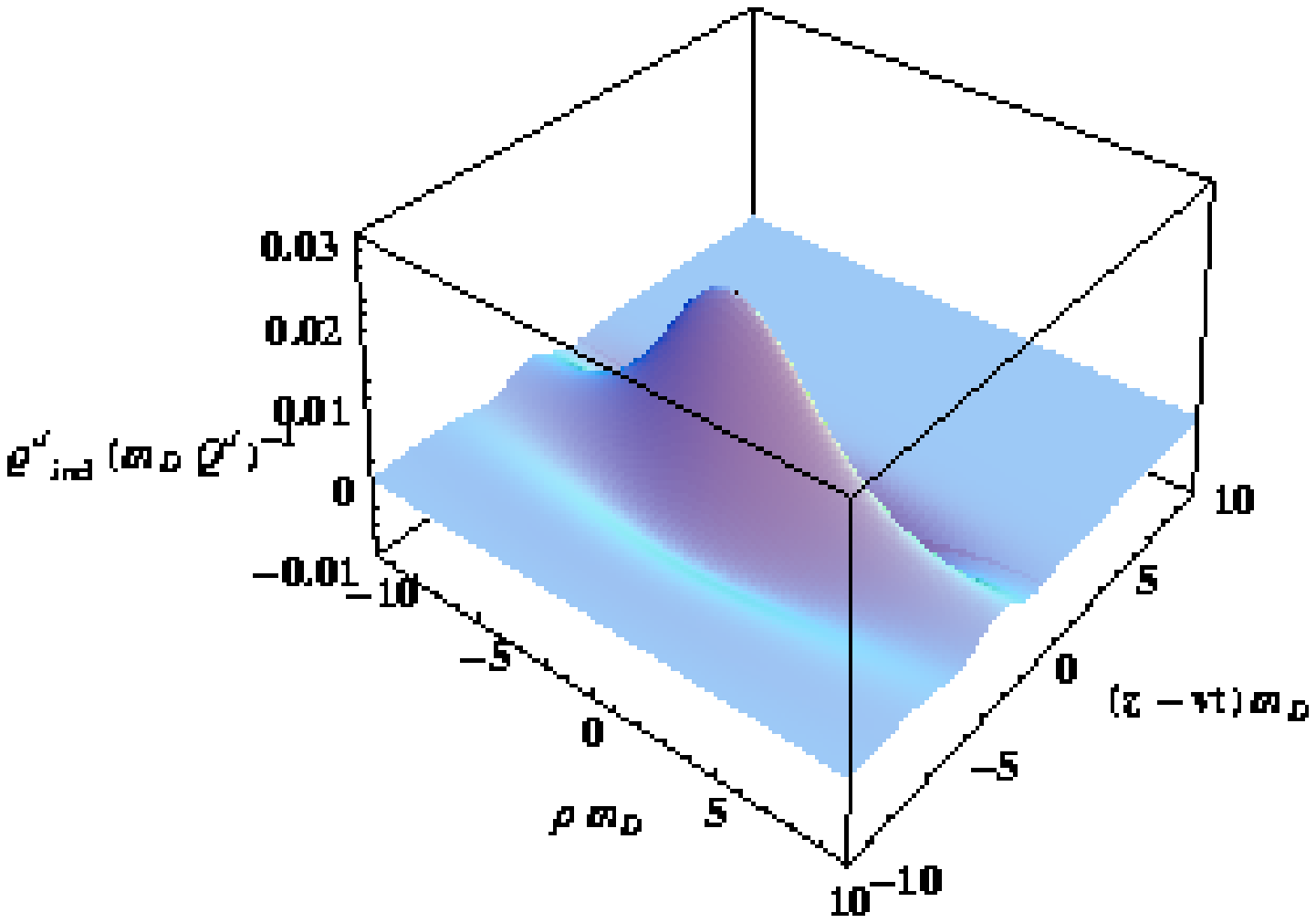}
\end{minipage}
\hfill
\begin{minipage}[t]{8cm}
\includegraphics[width=8cm,keepaspectratio]{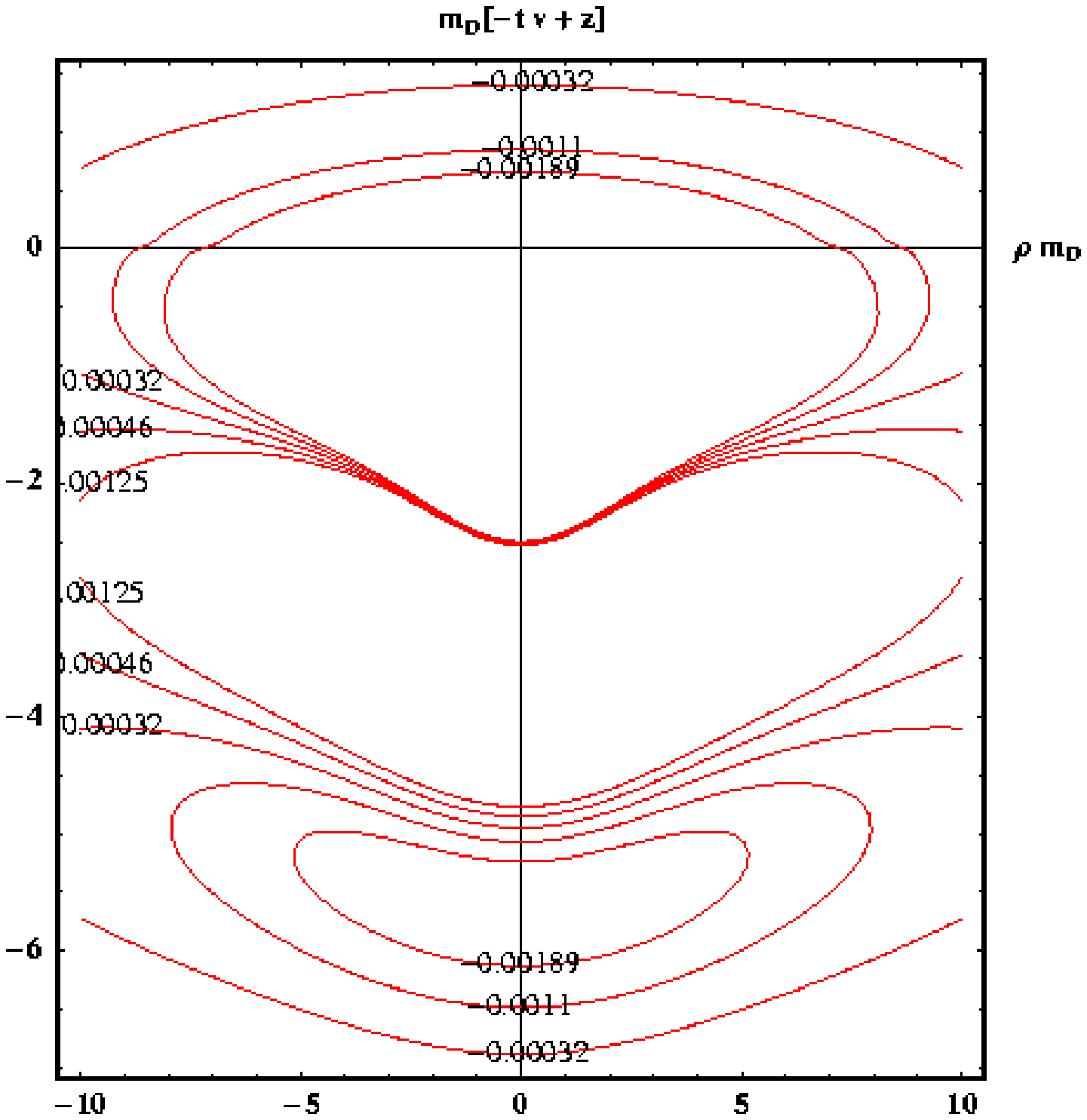}
\end{minipage}
\hfill
\begin{minipage}[t]{8cm}
\includegraphics[width=8cm,keepaspectratio]{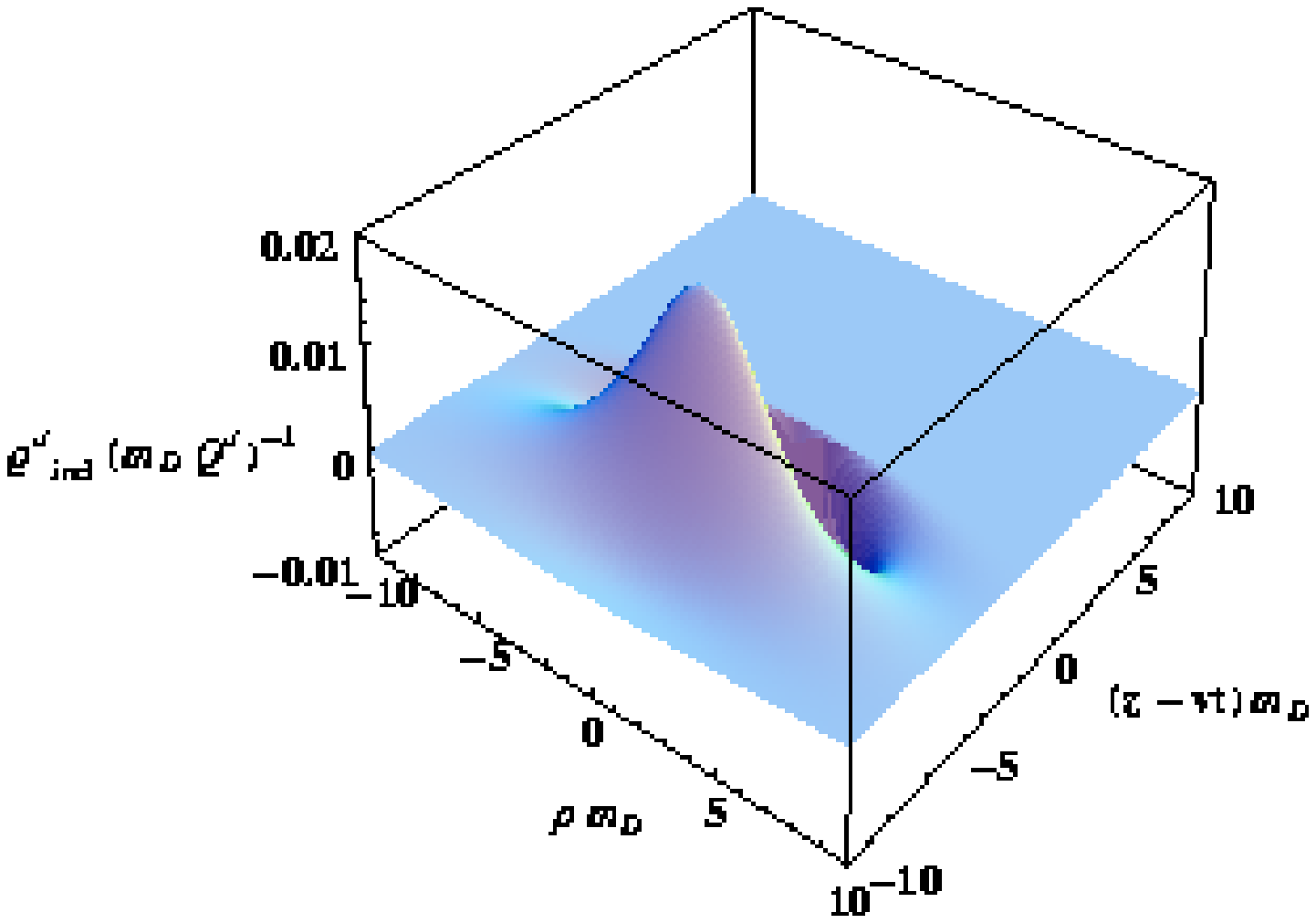}
\end{minipage}
\hfill
\begin{minipage}[t]{8cm}
\includegraphics[width=8cm,keepaspectratio]{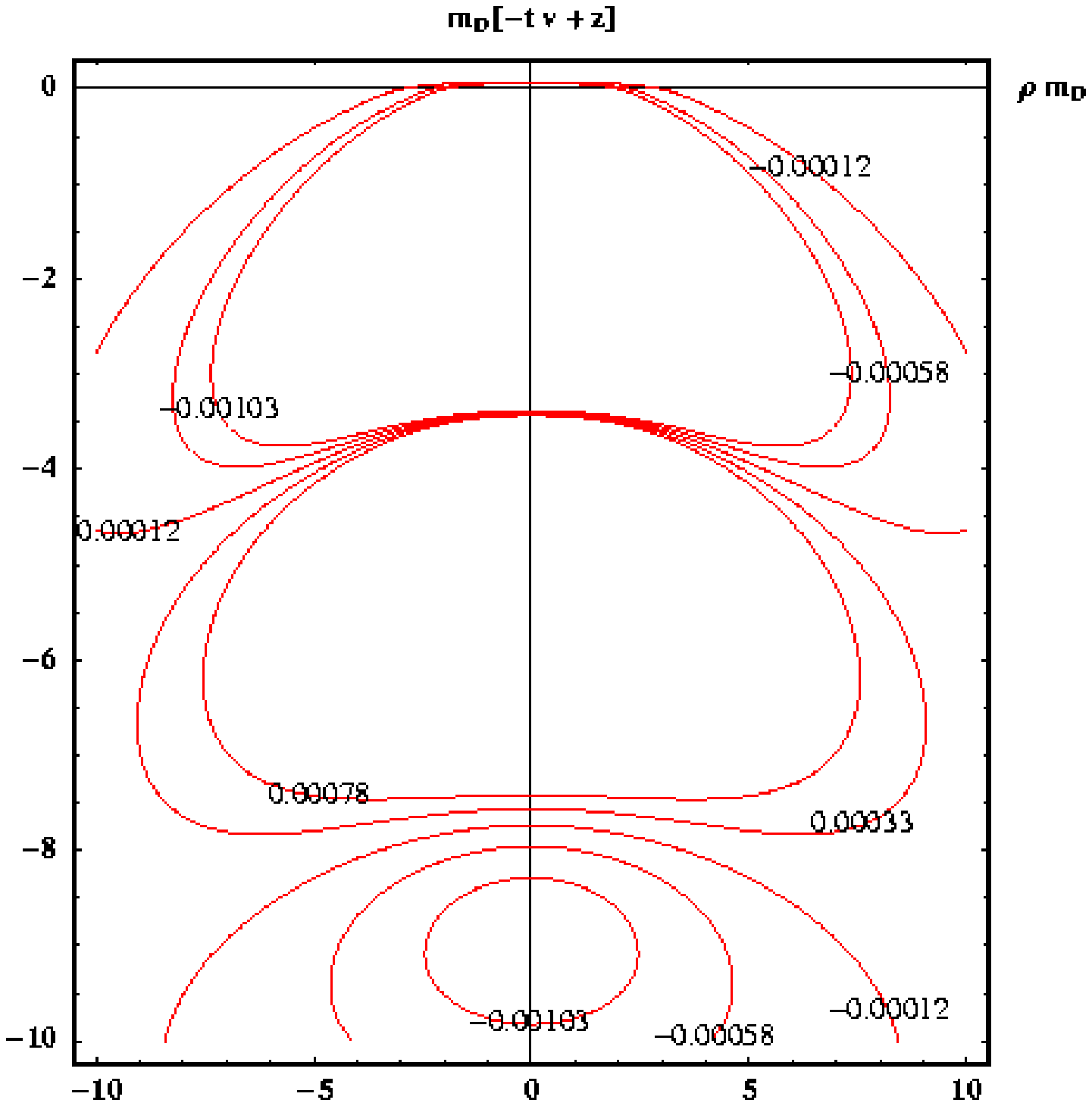}
\end{minipage}
\hfill
\begin{minipage}[t]{8cm}
\includegraphics[width=8cm,keepaspectratio]{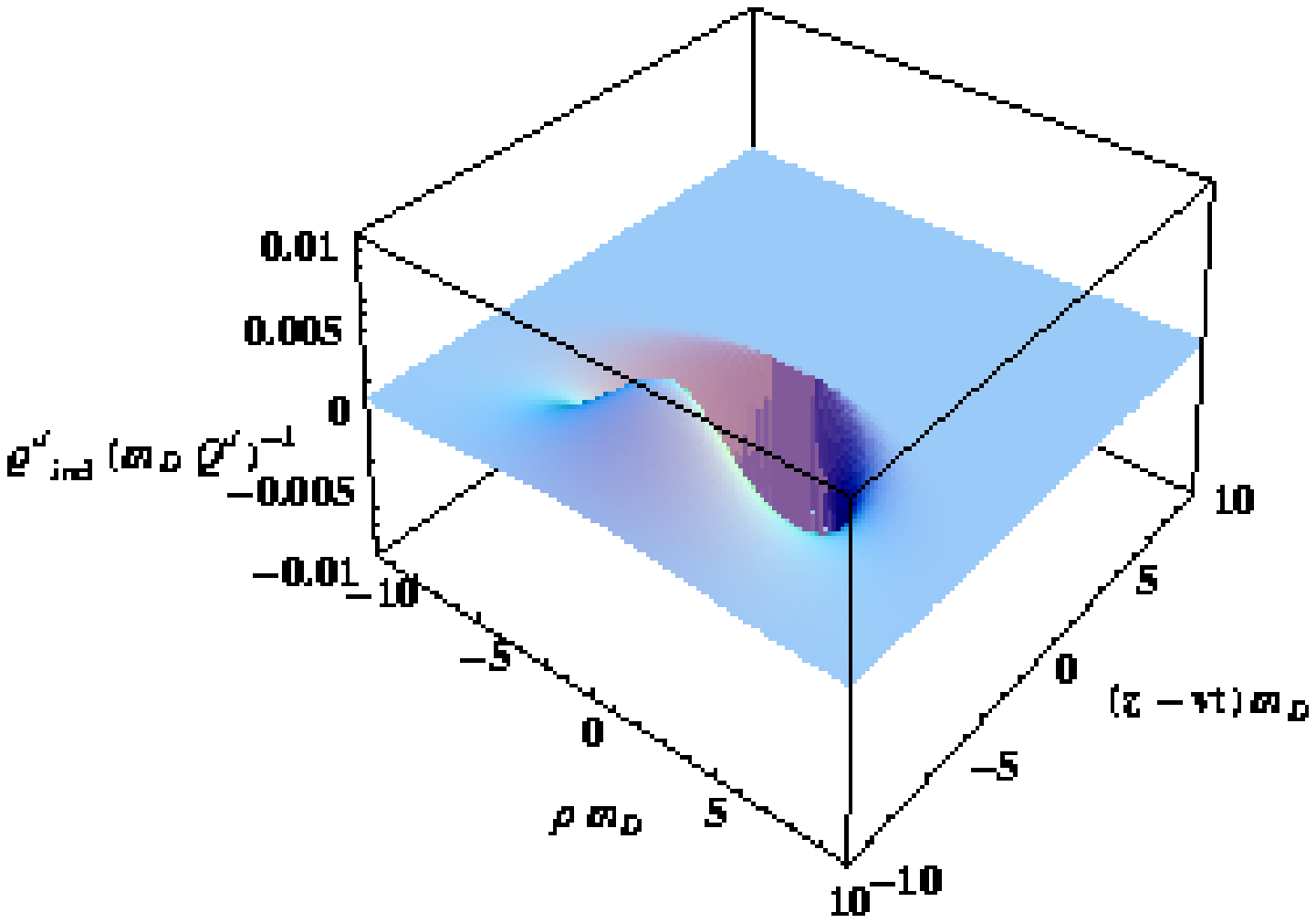}
\end{minipage}
\hfill
\begin{minipage}[t]{8cm}
\includegraphics[width=8cm,keepaspectratio]{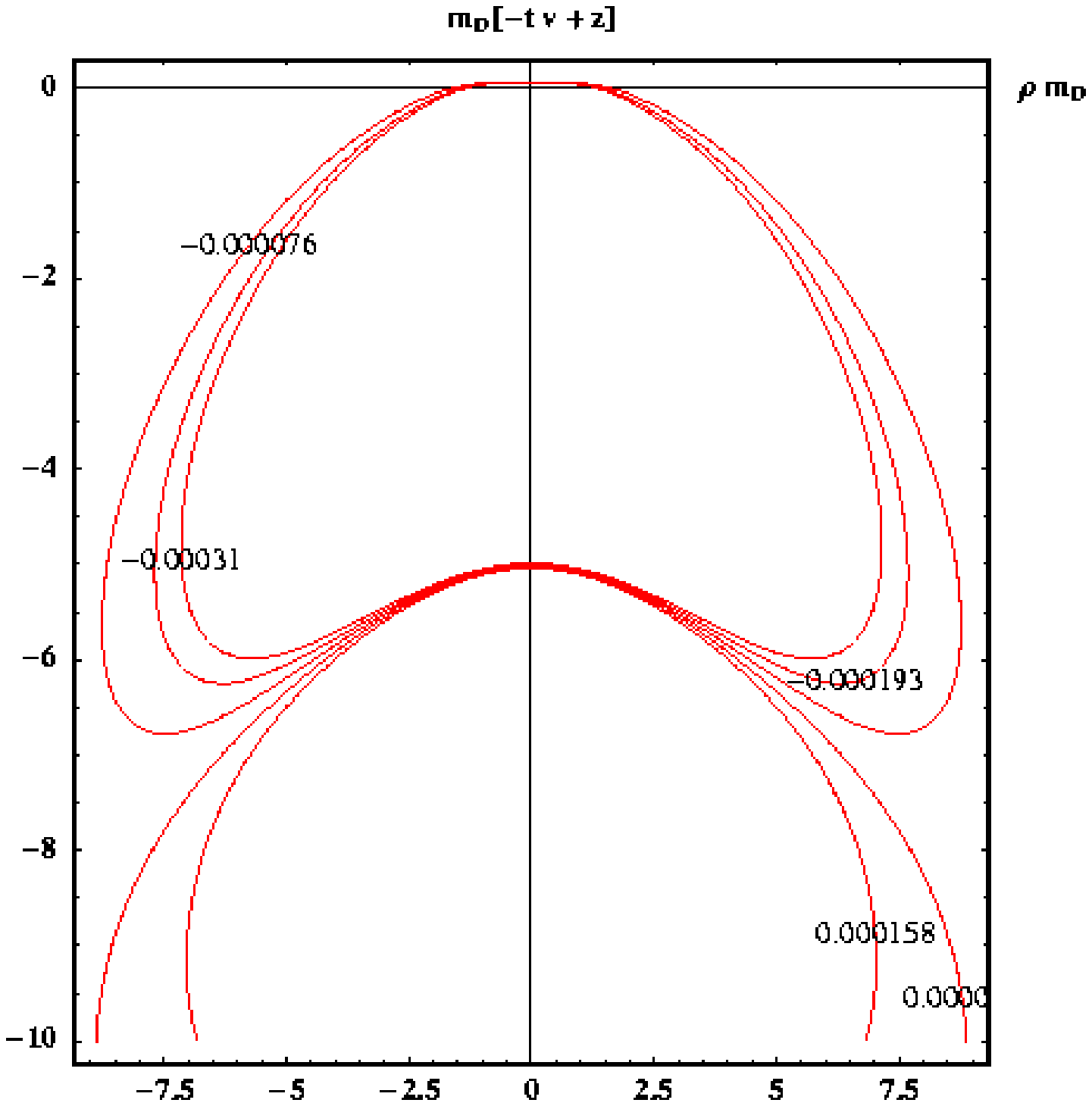}
\end{minipage}
\end{figure}

\begin{figure}[!htbp]
\begin{minipage}[t]{8cm}
\includegraphics[width=8cm,keepaspectratio]{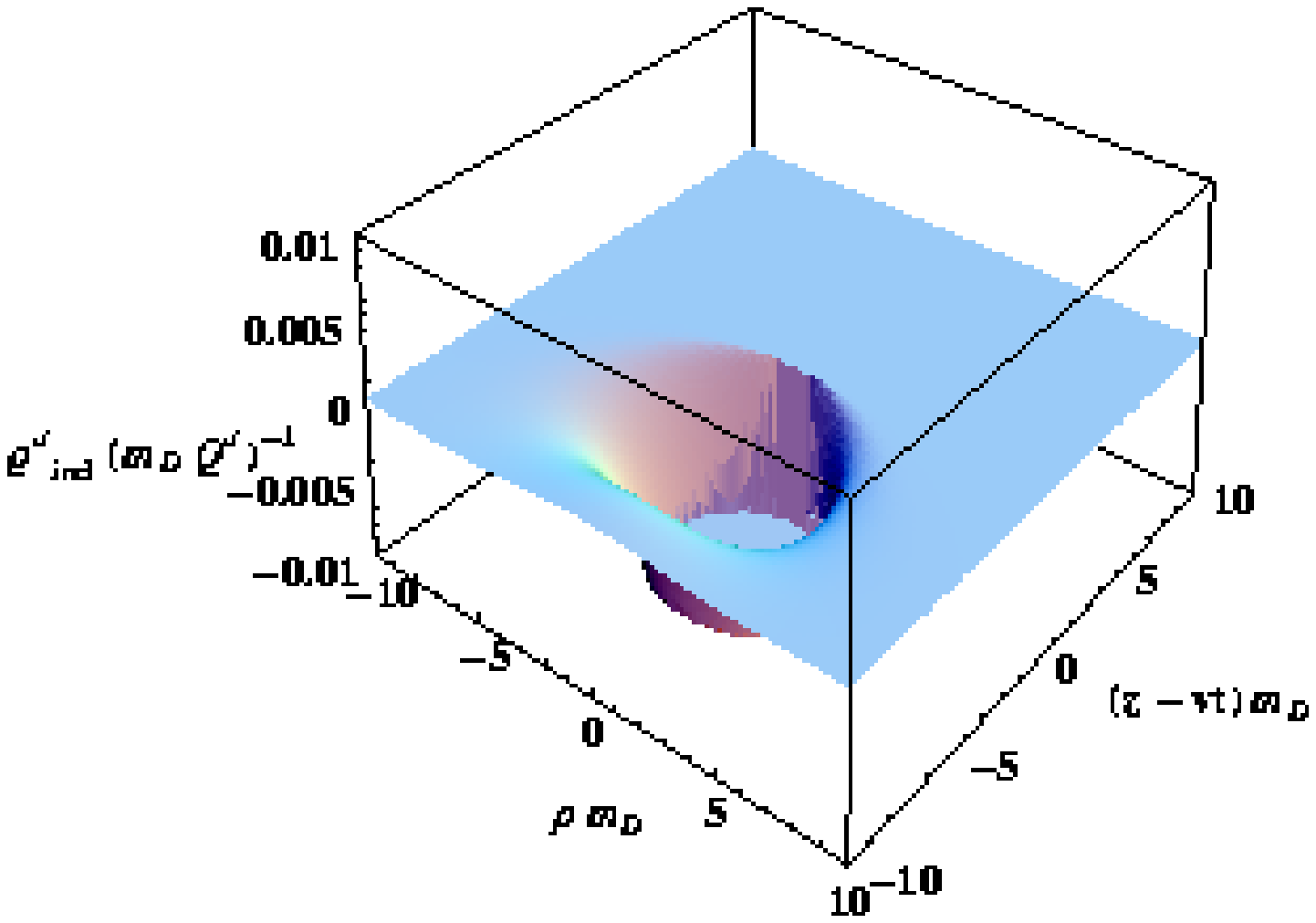}
\end{minipage}
\hfill
\begin{minipage}[t]{8cm}
\includegraphics[width=8cm,keepaspectratio]{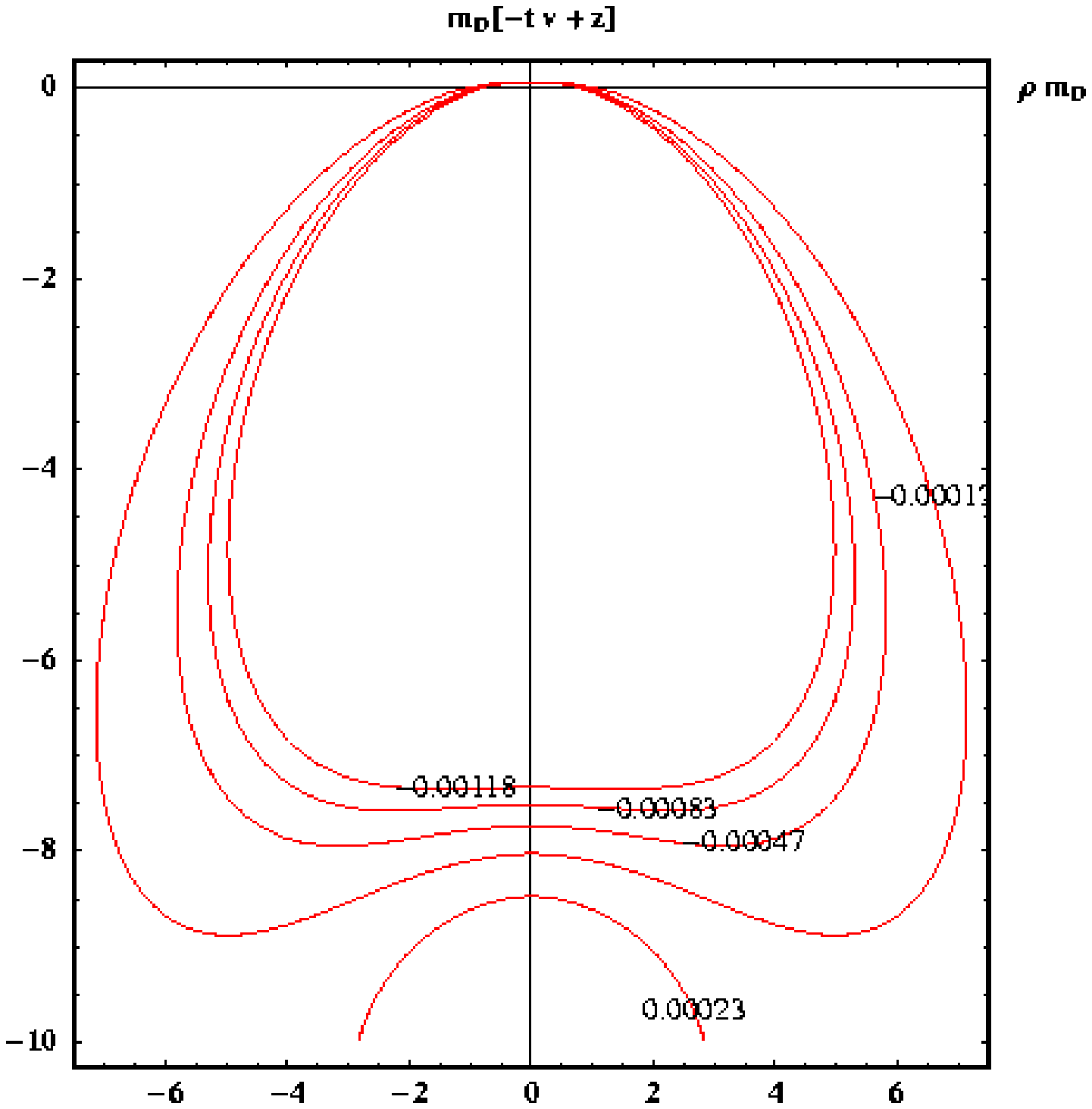}
\end{minipage}
\caption{\label{ichargedendity_v99} Left panel: Spatial distribution of 
the induced charge density for a parton moving with velocity $v=.99c$ 
with collisions for $(\nu=0, \, \,  0.2m_{D}, \, \, 0.5m_{D}, 0.8m_{D})$ (Top to Bottom).
Right panel: These plots show the corresponding contour plots}
\end{figure}

\newpage

\begin{figure}[!htbp]
\begin{minipage}[t]{8cm}
\includegraphics[width=8cm,keepaspectratio]{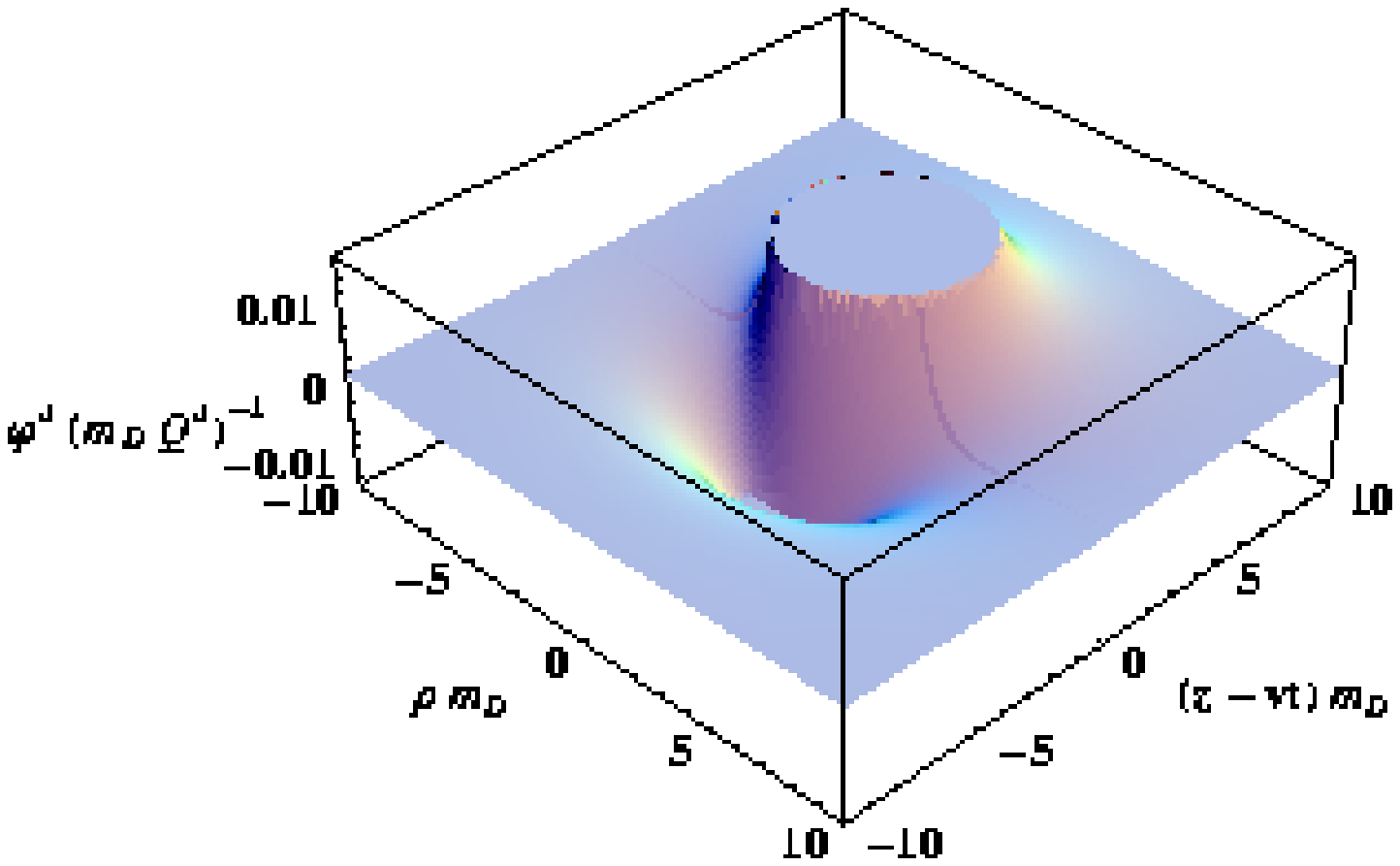}
\end{minipage}
\hfill
\begin{minipage}[t]{8cm}
\includegraphics[width=8cm,keepaspectratio]{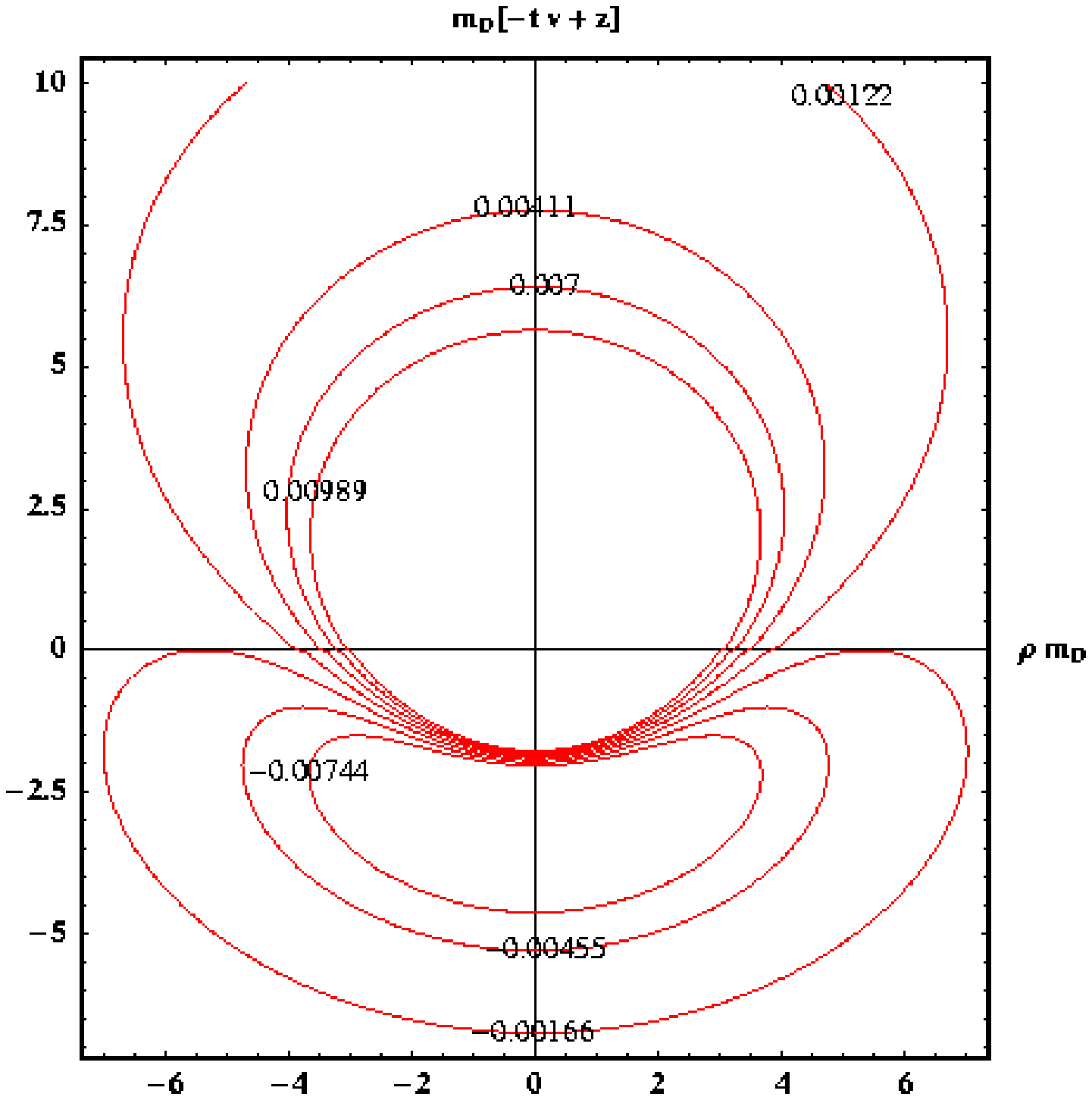}
\end{minipage}
\begin{minipage}[t]{8cm}
\includegraphics[width=8cm,keepaspectratio]{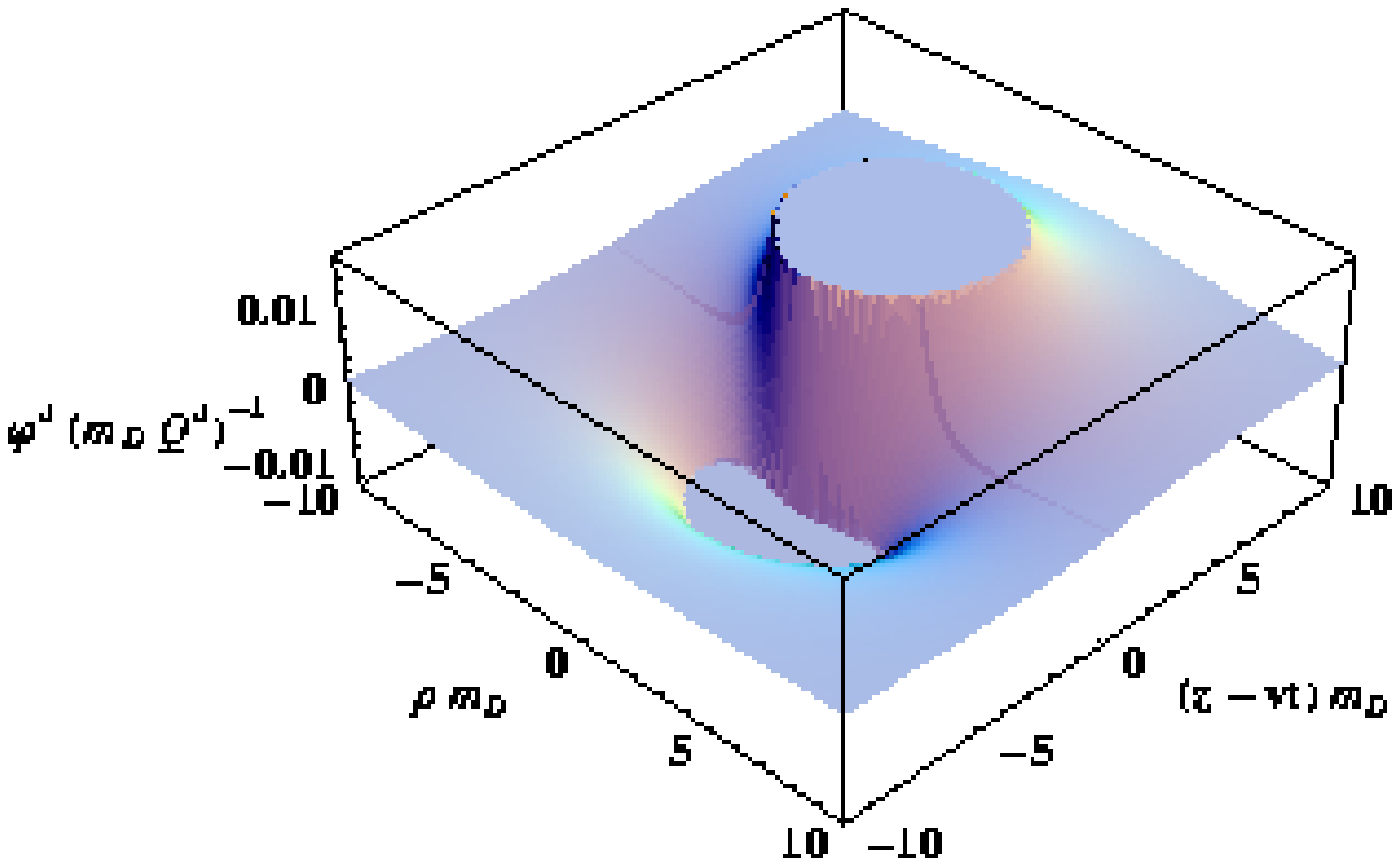}
\end{minipage}
\hfill
\begin{minipage}[t]{8cm}
\includegraphics[width=8cm,keepaspectratio]{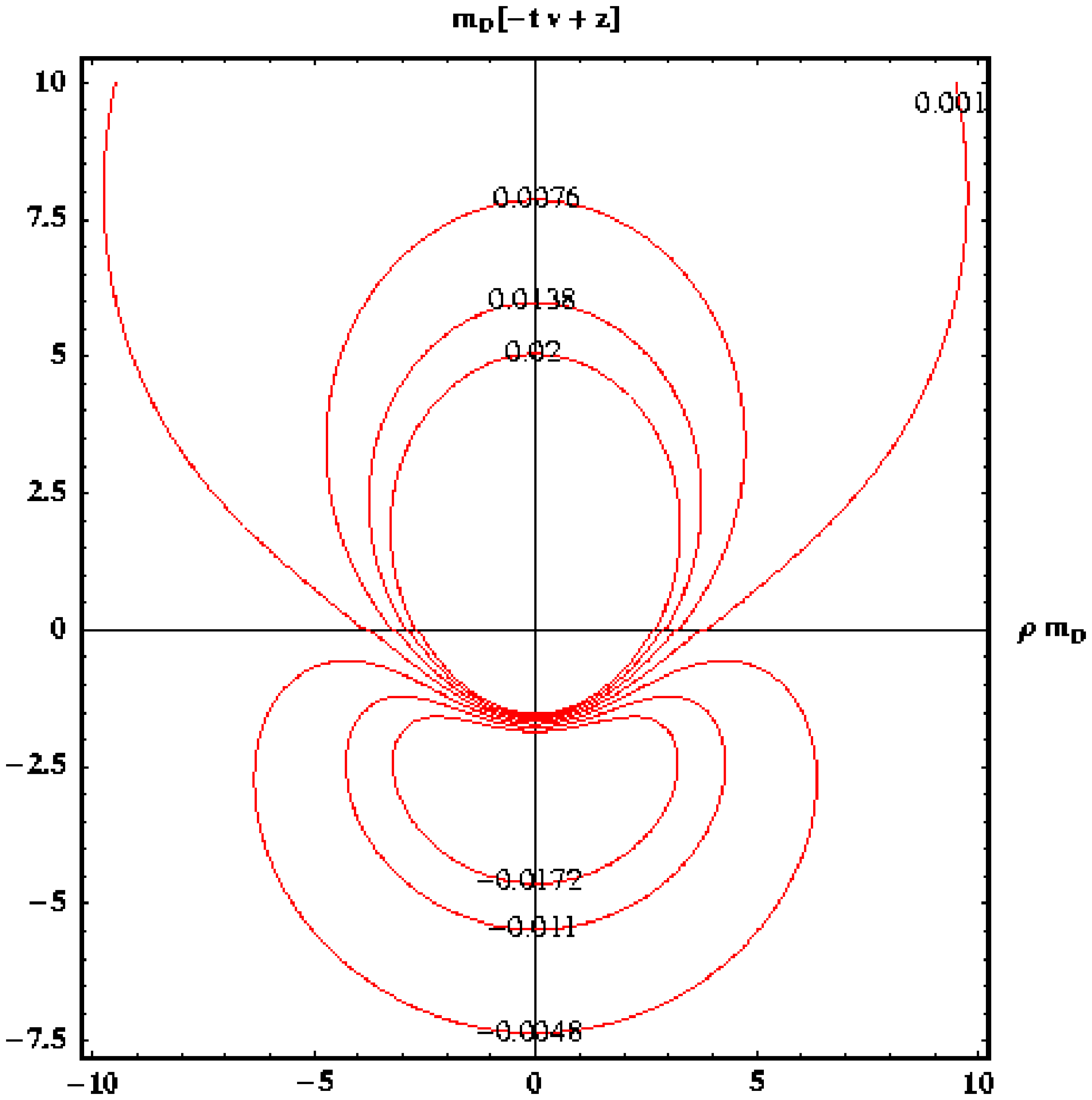}
\end{minipage}
\begin{minipage}[t]{8cm}
\includegraphics[width=8cm,keepaspectratio]{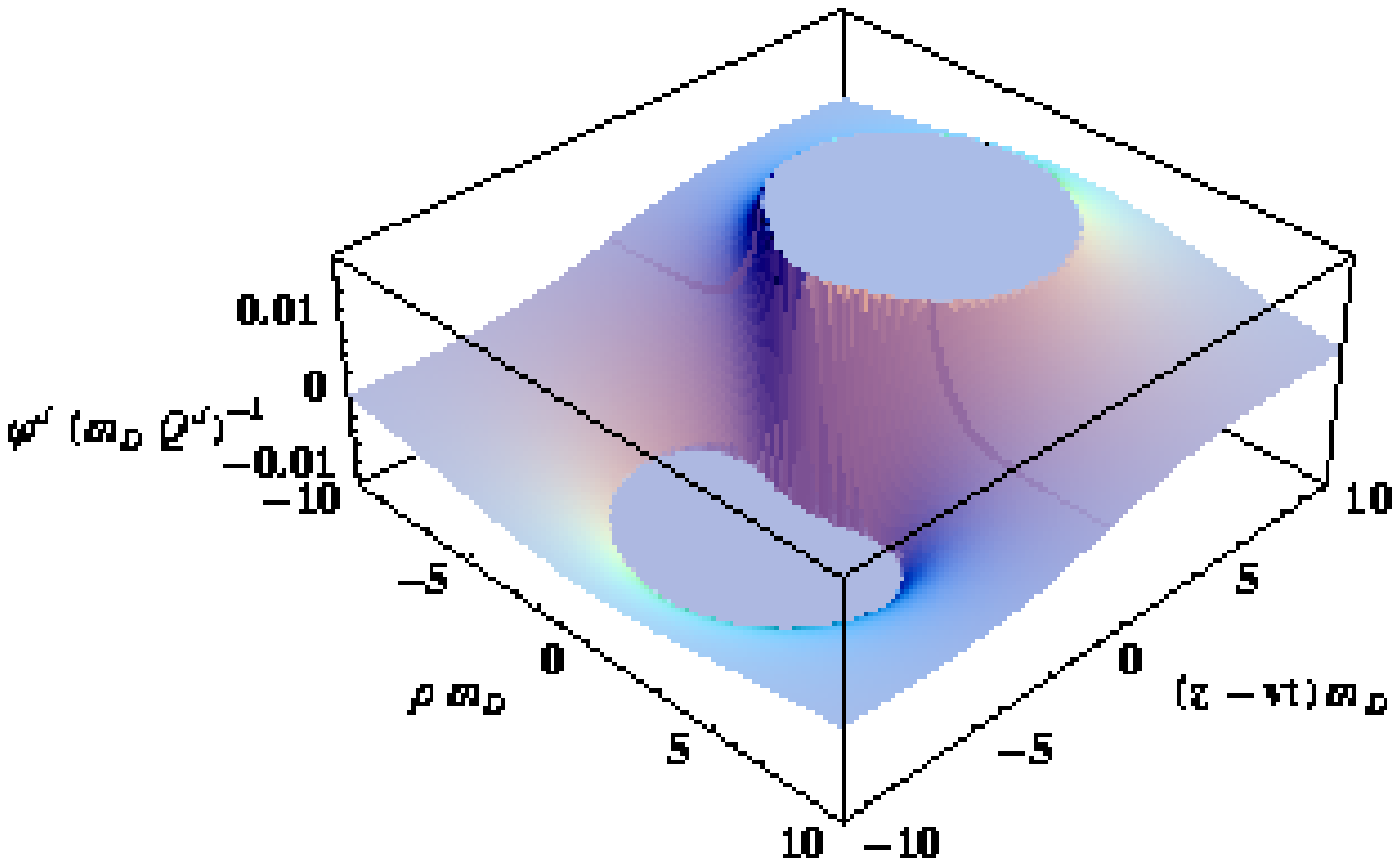}
\end{minipage}
\hfill
\begin{minipage}[t]{8cm}
\includegraphics[width=8cm,keepaspectratio]{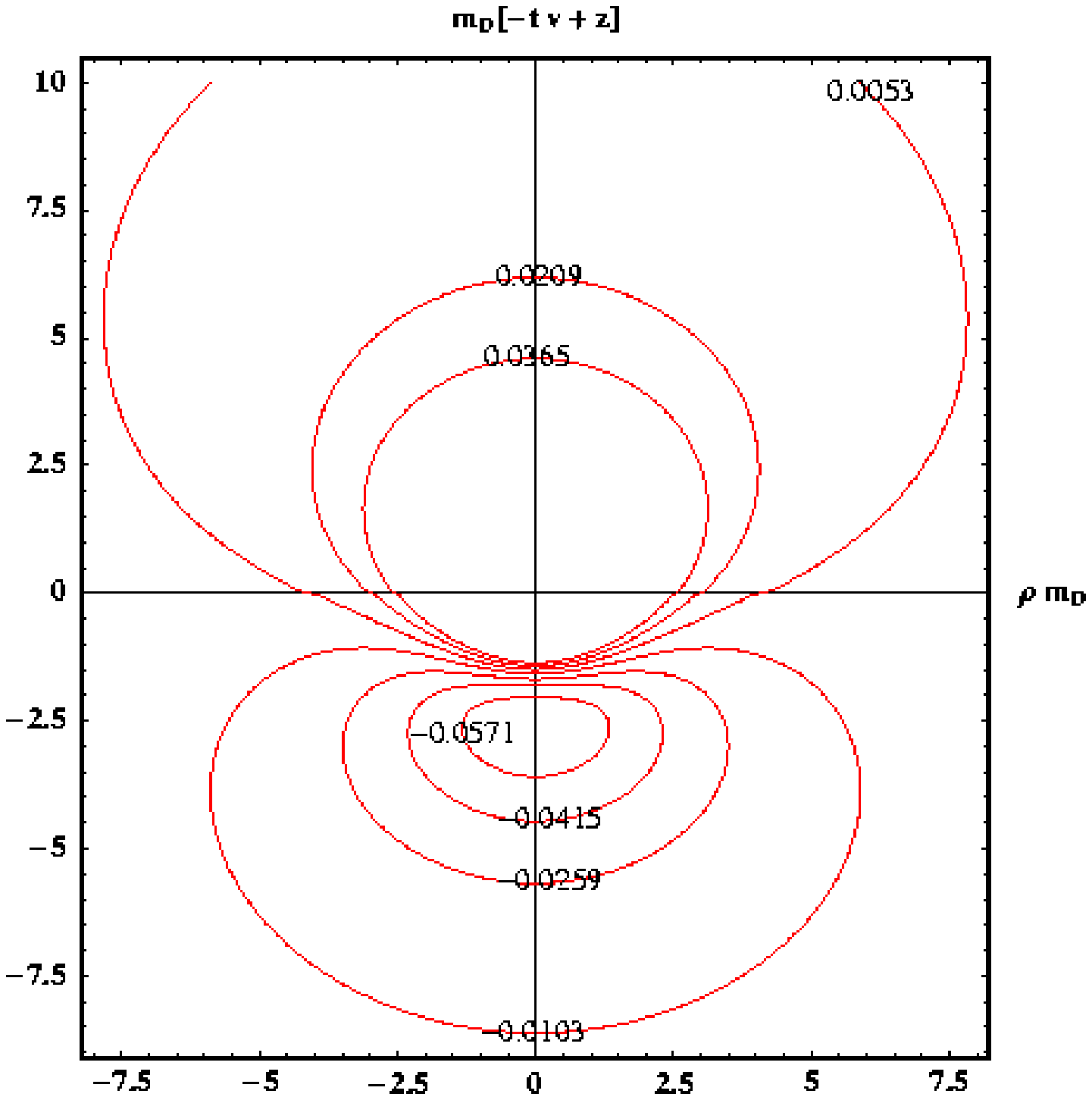}
\end{minipage}
\end{figure}
\begin{figure}[!htbp]
\begin{minipage}[t]{8cm}
\includegraphics[width=8cm,keepaspectratio]{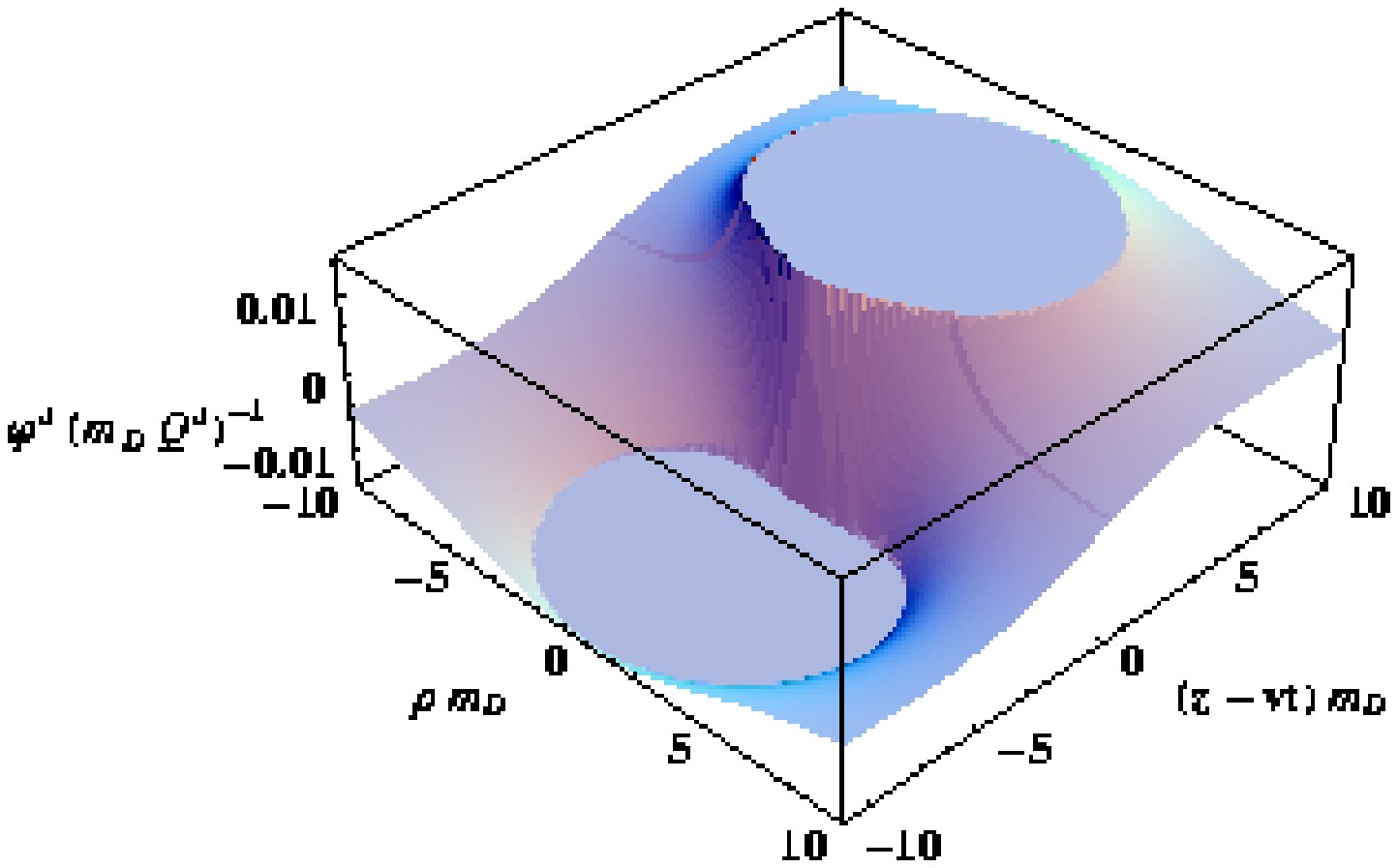}
\end{minipage}
\hfill
\begin{minipage}[t]{8cm}
\includegraphics[width=8cm,keepaspectratio]{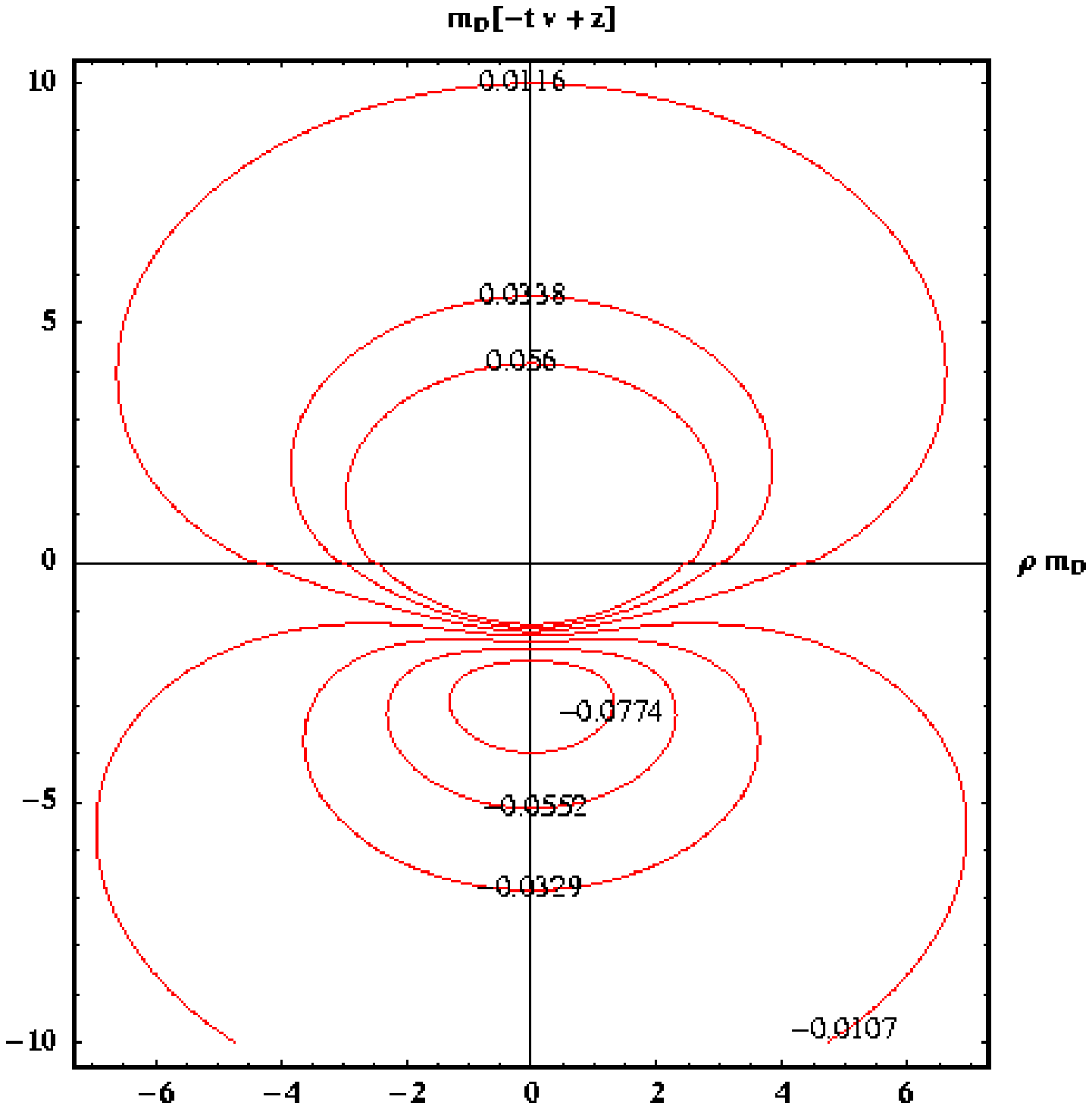}
\end{minipage}
\caption{\label{ipot_v55}  Left panel: Spatial distribution of the 
scaled wake potentials for a parton moving with velocity $v=.55c$ 
with collisions for $(\nu=0, \, \, 0.2m_{D}, \, \, 0.5m_{D}, \, \, 
0.8m_{D}$ (top to bottom panel). Right panel: These plots show the 
corresponding equipotential lines.}
\end{figure}

\newpage

\begin{figure}[!htbp]
\begin{minipage}[t]{8cm}
\includegraphics[width=8cm,keepaspectratio]{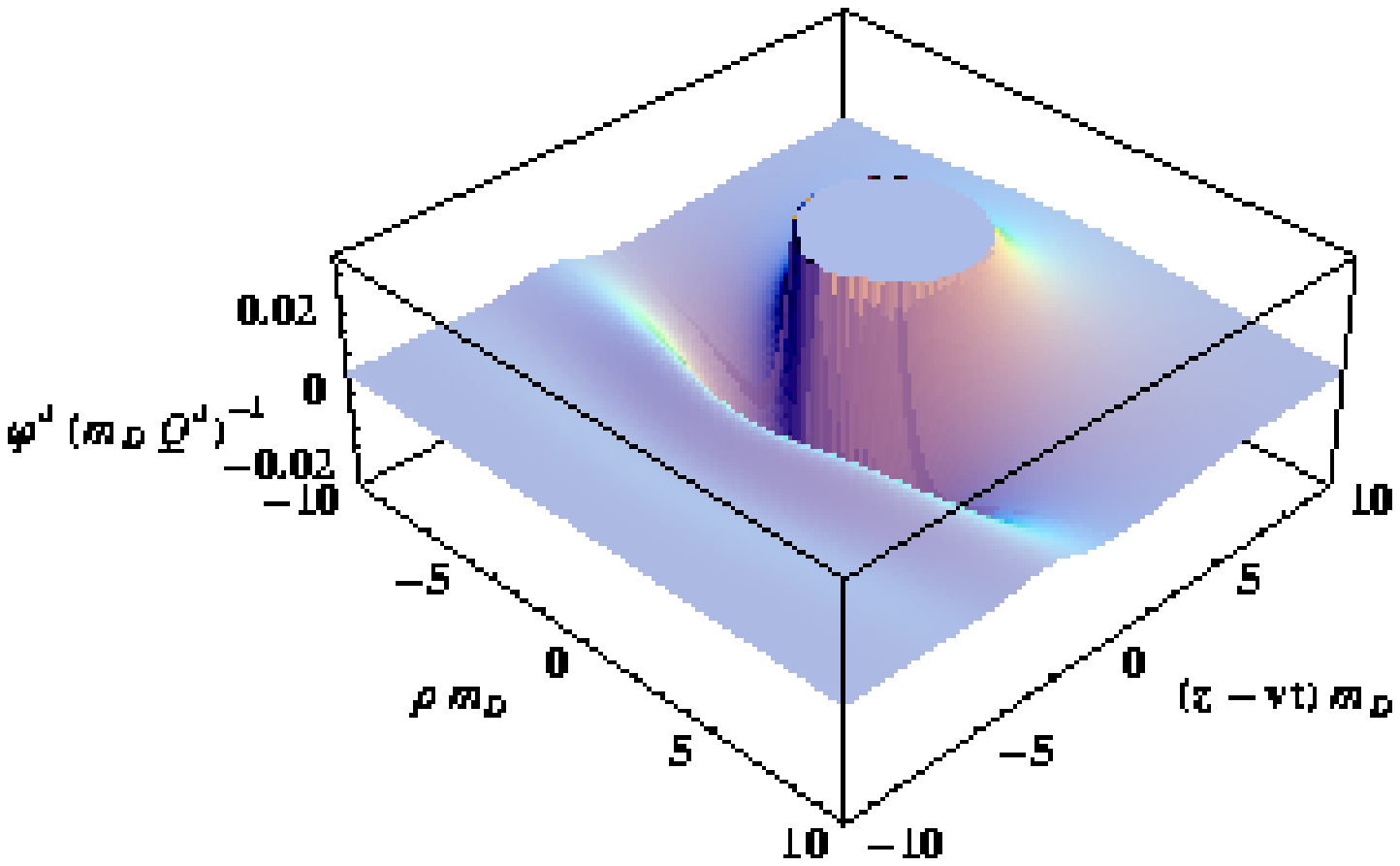}
\end{minipage}
\hfill
\begin{minipage}[t]{8cm}
\includegraphics[width=8cm,keepaspectratio]{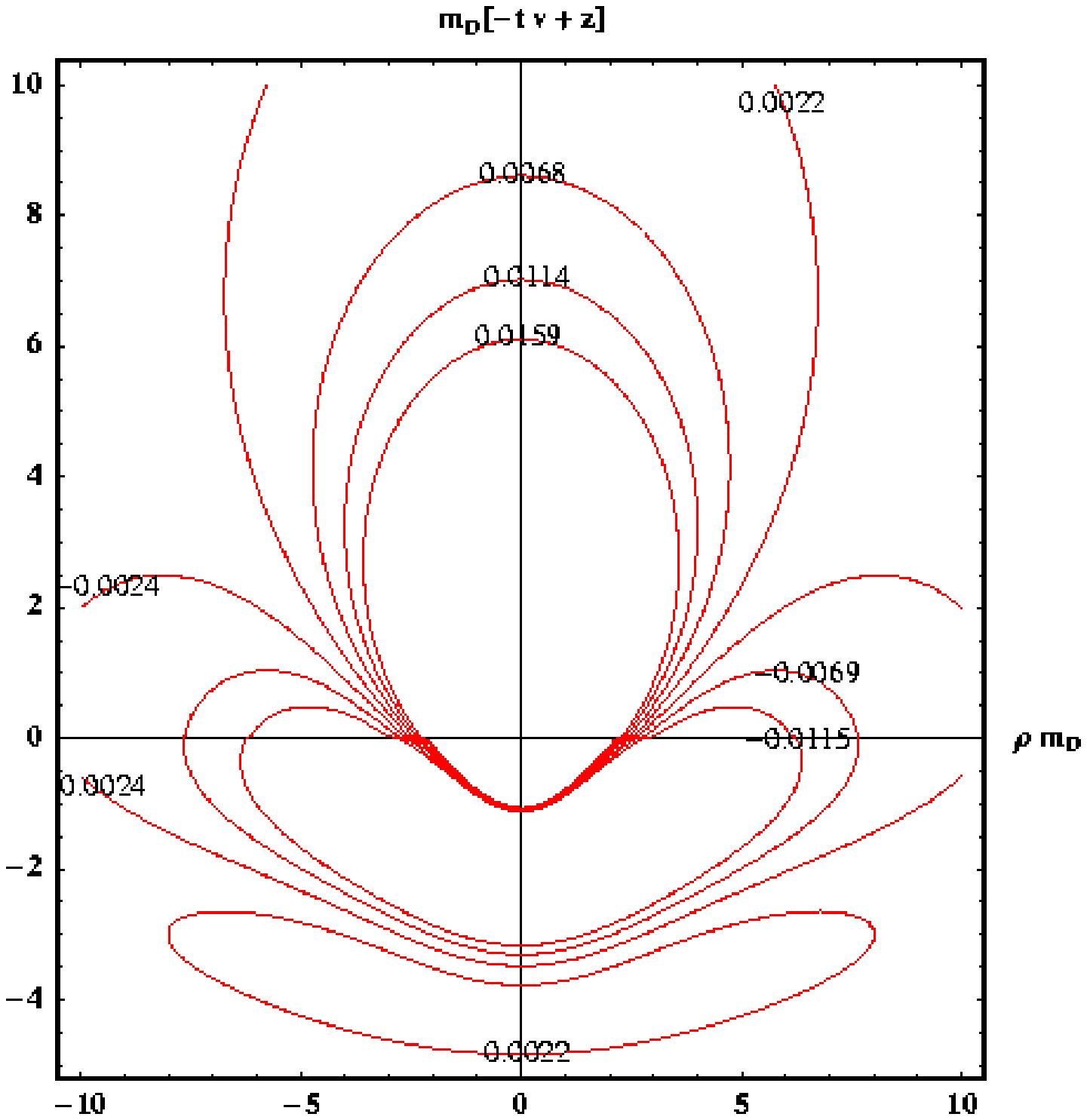}
\end{minipage}
\begin{minipage}[t]{8cm}
\includegraphics[width=8cm,keepaspectratio]{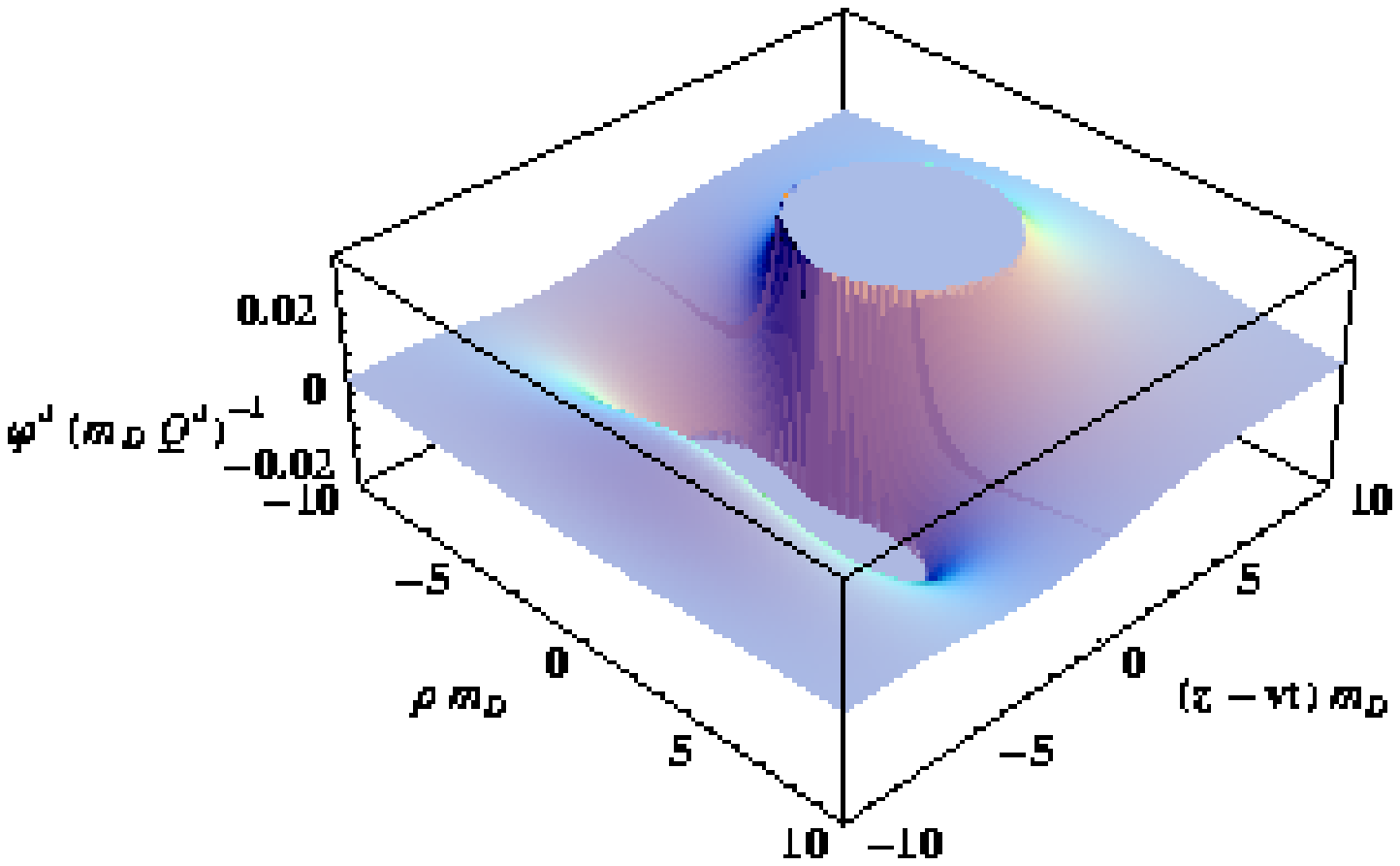}
\end{minipage}
\hfill
\begin{minipage}[t]{8cm}
\includegraphics[width=8cm,keepaspectratio]{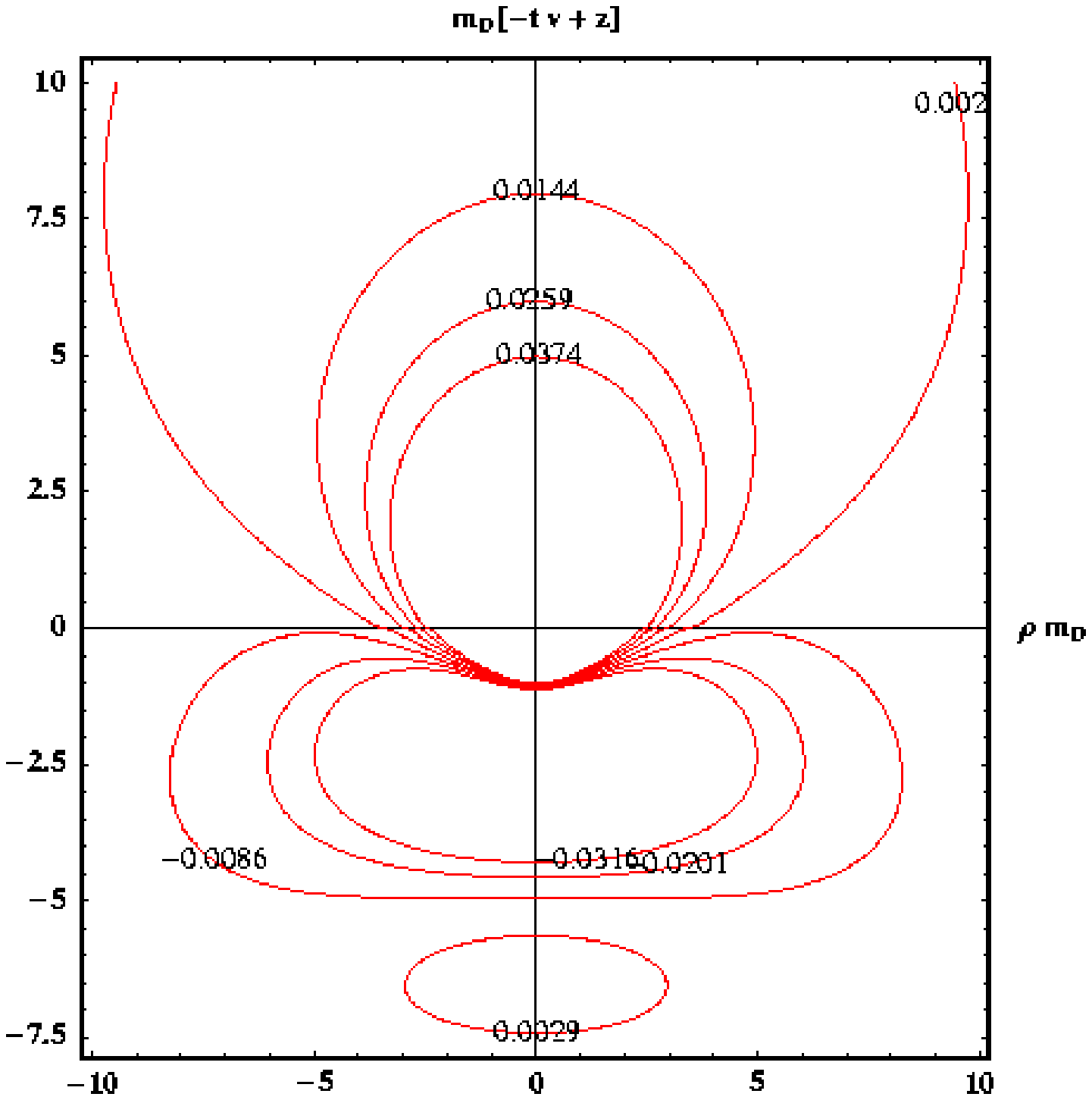}
\end{minipage}
\begin{minipage}[t]{8cm}
\includegraphics[width=8cm,keepaspectratio]{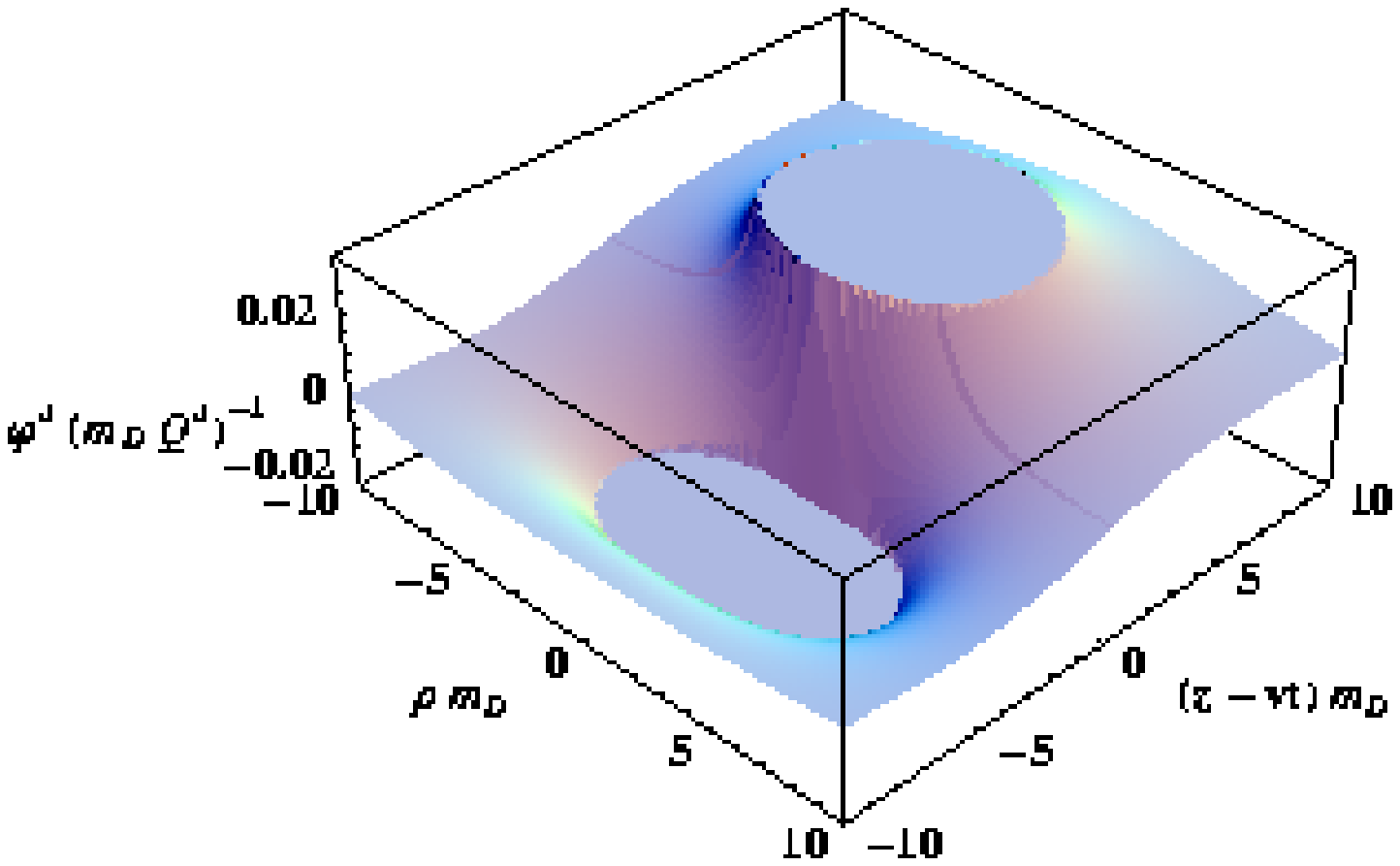}
\end{minipage}
\hfill
\begin{minipage}[t]{8cm}
\includegraphics[width=8cm,keepaspectratio]{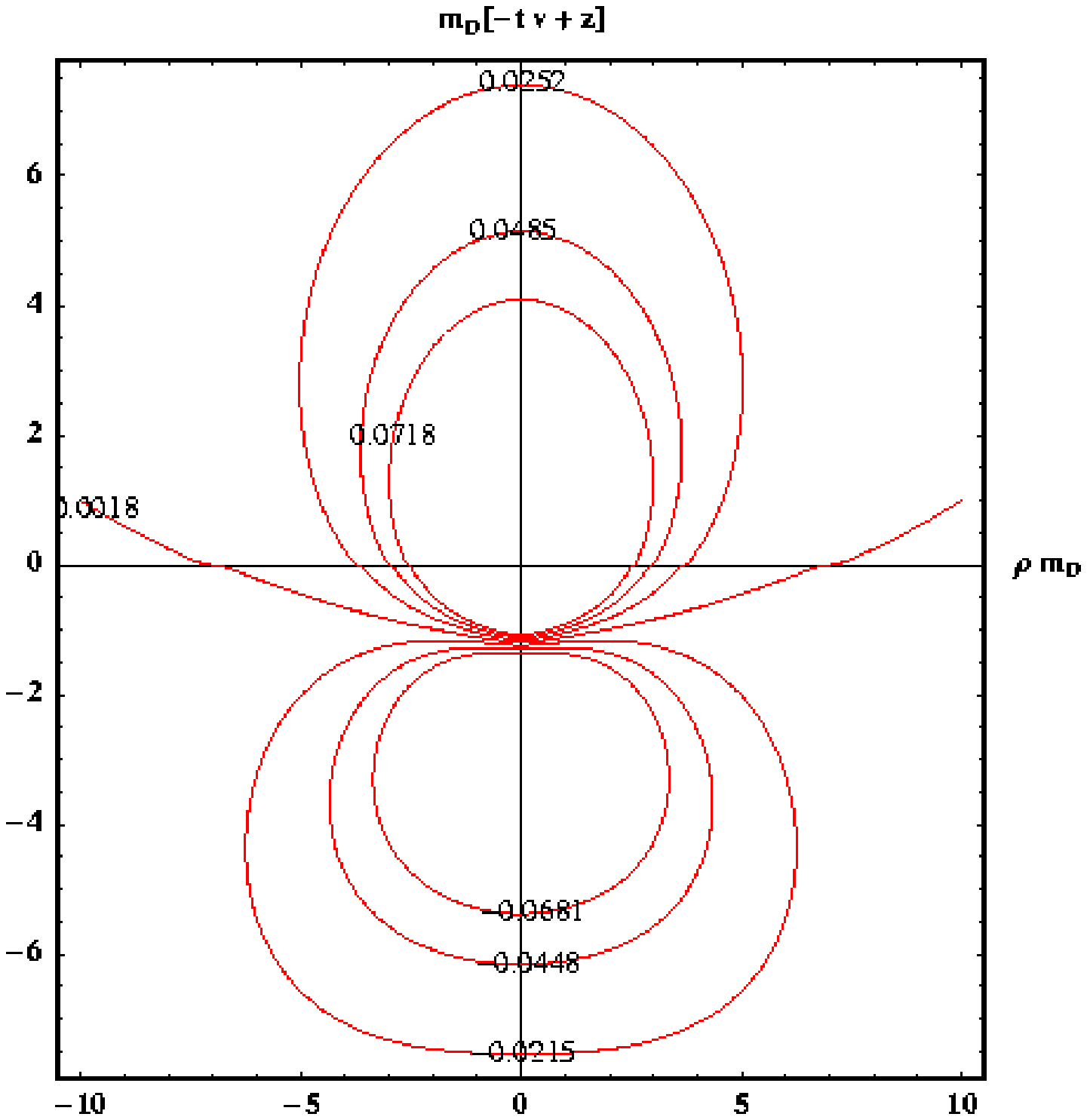}
\end{minipage}
\end{figure}
\begin{figure}[!htbp]
\begin{minipage}[t]{8cm}
\includegraphics[width=8cm,keepaspectratio]{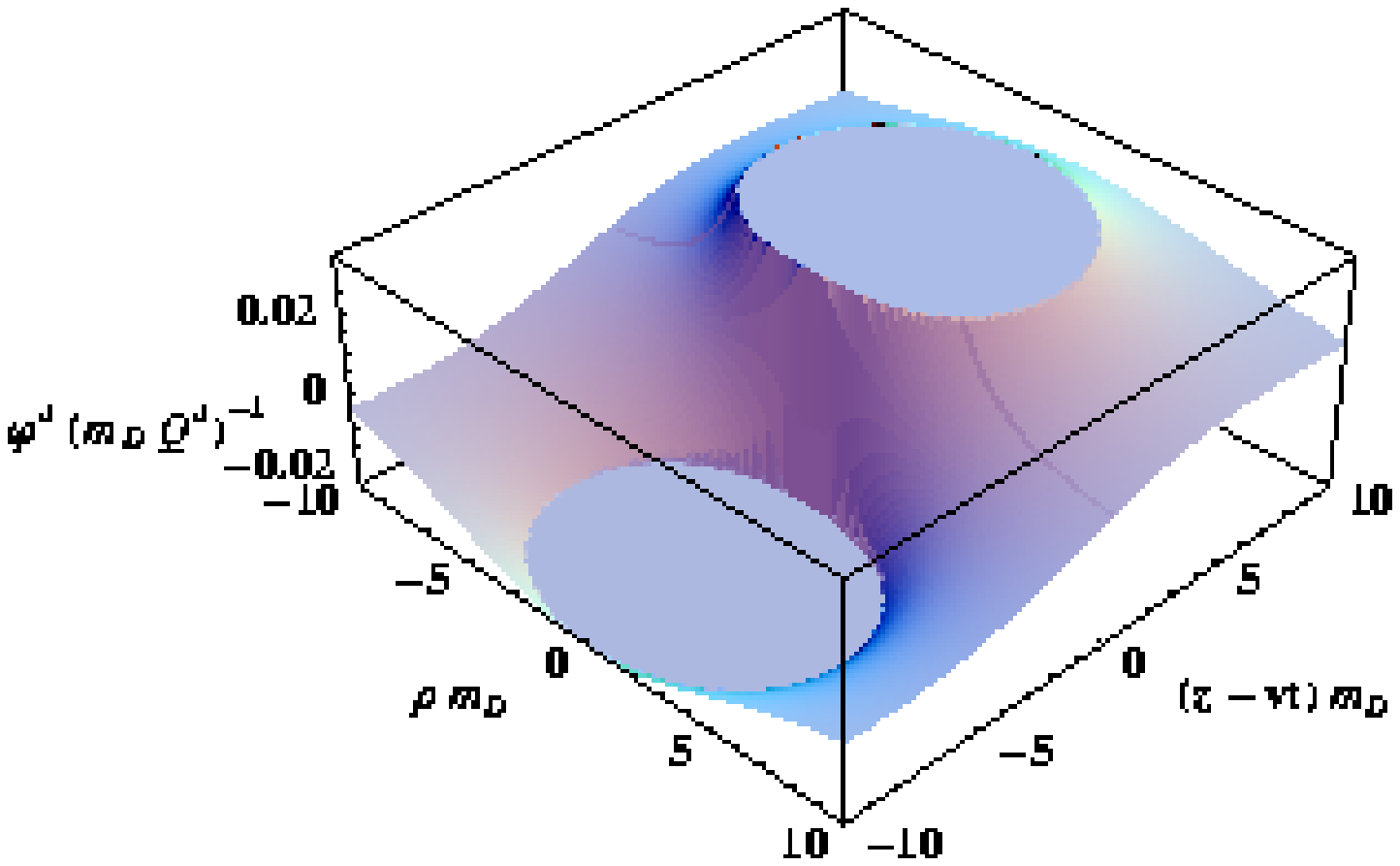}
\end{minipage}
\hfill
\begin{minipage}[t]{8cm}
\includegraphics[width=8cm,keepaspectratio]{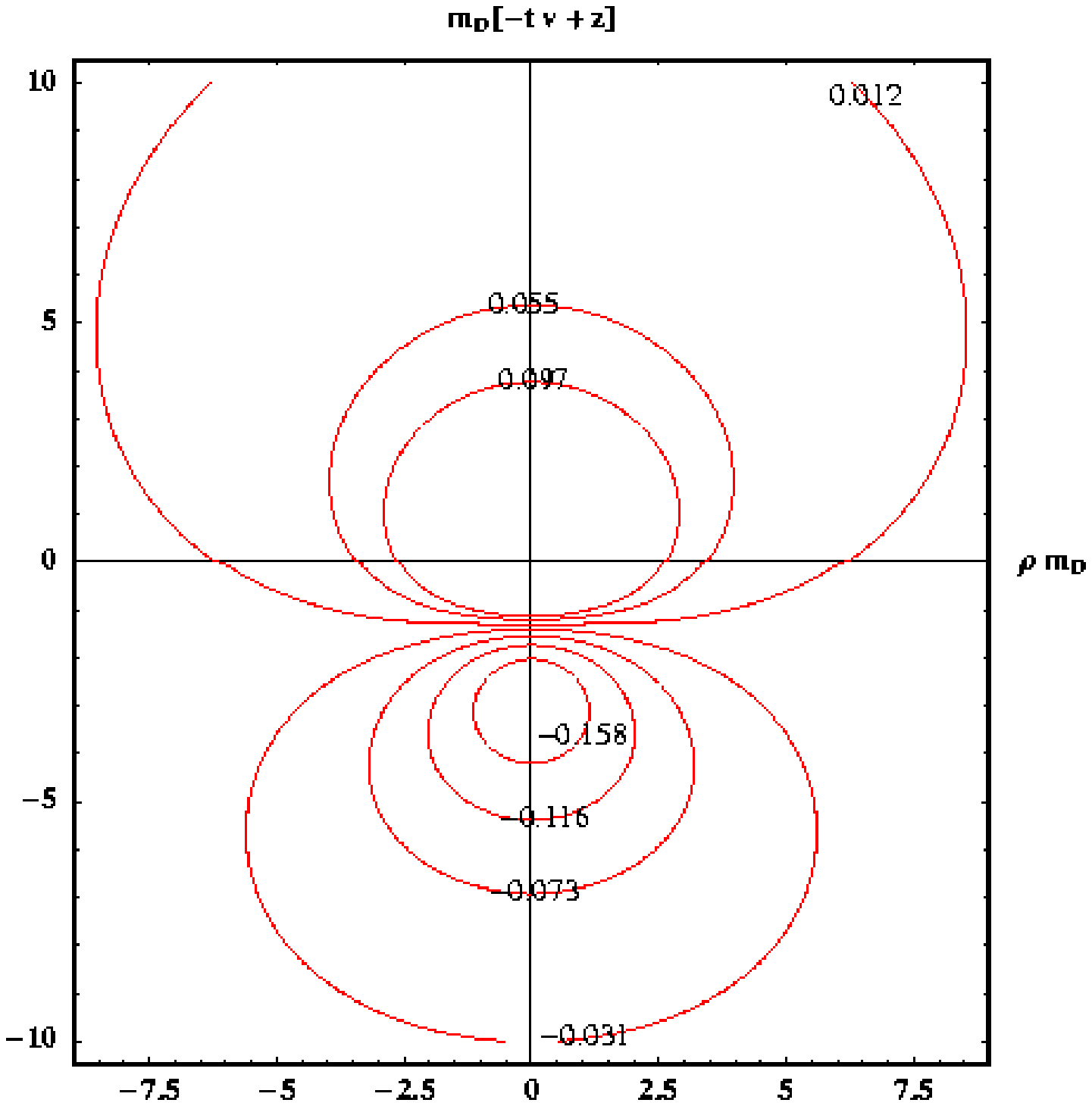}
\end{minipage}
\caption{\label{ipot_v99}  Left panel: Spatial distribution of the 
scaled wake potentials for a parton moving with velocity $v=.99c$ 
with collisions for $(\nu=0, \, \, 0.2m_{D}, \, \, 0.5m_{D}, \, \, 
0.8m_{D}$ (top to bottom panel). Right panel: These plots show 
the corresponding equipotential lines.}
\end{figure}


\begin{thebibliography}{99}
\bibitem{remark1}
The wakes discussed here are associated with the collisional energy loss only 
not with the radiative \cite{Baier2000}. However, the latter
might also produce a large-angle radiation of gluons in addition to the one
by Mach cones \cite{Vitev2005}.
\bibitem{remark2}
In Ref.\cite{Carrington2004} the BGK-approach has been derived for the QED case.
However, since the Vlasov aproach for QED and QCD is the same 
(apart from color and flavor factors), this approach can be used also for QCD
at least as a model for taking collisions into account (see also Ref.\cite{Schenke2006}).
\bibitem{remark4}
We restrict to the physical Riemann sheet, i.e. $\ln{z}=\ln |z| + i\theta + 2 i n \pi$ 
with $n=0$.
\end{thebibliography}
\end{document}